\documentclass[preprint,rmp,aps]{revtex4}

\begin{document}

\bibliographystyle{prsty}



\title{ Laws in Darwinian Evolutionary Theory }
\author{ P. Ao }
\address{ Department of Mechanical Engineering,
          University of Washington, Seattle, WA 98195, USA }


\begin{abstract}
In the present article the recent works to formulate laws in
Darwinian evolutionary dynamics are discussed. Although there is a
strong consensus that general laws in biology may exist, opinions
opposing such suggestion are abundant. Based on recent progress in
both mathematics and biology, another attempt to address this
issue is made in the present article. Specifically, three laws
which form a mathematical framework for the evolutionary dynamics
in biology are postulated. The second law is most quantitative and
is explicitly expressed in the unique form of a stochastic
differential equation. Salient features of Darwinian evolutionary
dynamics are captured by this law: the probabilistic nature of
evolution, ascendancy, and the adaptive landscape. Four dynamical
elements are introduced in this formulation: the ascendant matrix,
the transverse matrix, the Wright evolutionary potential, and the
stochastic drive. The first law may be regarded as a special case
of the second law. It gives the reference point to discuss the
evolutionary dynamics. The third law describes the relationship
between the focused level of description to its lower and higher
ones, and defines the dichotomy of deterministic and stochastic
drives. It is an acknowledgement of the hierarchical structure in
biology. A new interpretation of Fisher's fundamental theorem of
natural selection is provided in terms of the F-Theorem. The
proposed laws are based on continuous representation in both time
and population. Their generic nature is demonstrated through their
equivalence to classical formulations. The present three laws
appear to provide a coherent framework for the further development
of the subject.  \\
(  Published in {\bf Physics of Life Reviews}, {\bf 2} (2) June
{\bf 2005}, pp117-156 )
\end{abstract}



\maketitle

\tableofcontents

{\ }

{\ }

{\ }

\noindent {{\it {\large Nothing in Biology Makes Sense Except in
the Light of Evolution } }}

 {\ } {\ } {\ } {\ } {\ } {\ } {\ } {\ } {\ } {\ } {\ } {\ } {\ } {\ } {\ } {\ } {\
 } {\ } {\ } {\ } {\ } {\ } {\ } {\ } {\ } {\ }  {\ } {\ } {\ }
 {\bf Theodosius Dobzhansky (1900-1975)}

\section{ Introduction }

\subsection{Background}

Darwinian evolutionary theory occupies a unique position in
biological sciences. Progress in experimental biology and new data
from field observation after the neo-Darwinian synthesis pose new
questions to be answered and call attention to previously
unanswered questions. One of the central questions is whether or
not there exist general laws in biology. Biologists have been
responding to those demands with tremendous activities
\cite{burger,ddmt,ewens,felsenstein,gavrilets,gould,mayr2004,rice,turchin}.
There was a critical discussion \cite{grene} of existing theories.
Courageous exploration \cite{kauffman} of new theories was carried
out to probe the origin of order.  Serious consideration of the
epistasis of gene interactions \cite{epistasis} was made. Shifting
balance process \cite{coyne,goodnight,peck} was reevaluated. There
is a continuous interest in the concept of species
\cite{mayr2004,pigliucci}. The fundamental theorem of natural
selection \cite{crow,ewens,grafen} has been constantly reexamined.
Various elegant mathematical models of speciation
\cite{ddmt,stewart,gavrilets,kisdi} were explored. Even new
philosophical implications have been speculated \cite{es1998} and
discussed \cite{mayr2004}. Such efforts have enriched both
quantitative and conceptual understanding of evolution.

It has been noted \cite{epistasis} that a quantitative formulation
of evolutionary dynamics may be of most importance to answer the
new questions. Verbal description has been found to be inadequate,
because the various deterministic and stochastic contributions to
evolutionary dynamics, either independent or interactive, are
often of same order of magnitude. Also, interactions can be highly
nonlinear and cooperative behaviors are abundant. For example, it
was from this consideration that Stewart \cite{stewart} built his
model based on symmetry-breaking and Gavrilets \cite{gavrilets}
advanced his holey adaptive landscape model. The present review
develops further this line of research. We make an attempt to
formulate a general and quantitative mathematical framework which
appears broad enough to incorporate the ideas of nonlinear
symmetry-breaking \cite{stewart} and diffusion in the holey
adaptive landscape \cite{gavrilets}. It is conceptually consistent
with Darwinian dynamics. The formulation appears to capture the
dynamical part of the evolution by natural selection
\cite{darwin1858,darwin1958}. In particular, two of the most
influential quantitative concepts in evolutionary biology, the
adaptive landscape of Wright \cite{wright} and the fundamental
theorem of natural selection of Fisher \cite{fisher}, are built
naturally into the present formulation, after their reformulations
in the light of modern progress. The present work may also be
regarded as an attempt to unify approaches on nonequilibrium
processes from both biological and physical sciences.

\subsection{Language of Mathematical Description}

The mathematical approach in the present work is based on the
continuous representation which treats both time and state
variable (for example, populations of species in ecological
dynamics and frequencies of genotypes in population genetics) as
continuous variables. There are three reasons for choosing the
continuous representation. First, since the continuous
representation has been widely used in physical sciences, it would
be easy to make connections to those fields. Second, from a
mathematical point of view, any discrete process may be
represented by an appropriate continuous one. Hence there is no
loss in rigor to use continuous representation. Thirdly and most
importantly, the connection between discrete and continuous
representations has been well studied in biology, documented, for
example, in both a commentated collection of historical articles
\cite{li}, in a recent monograph \cite{burger} and in recent
textbooks \cite{ewens,rice}, and in an online book
\cite{felsenstein}, and has been successfully and widely employed
in population genetics. Those studies provide the needed link
between biological parameters appearing in real biological cases
and the corresponding continuous representation. Such connection
has also been studied in physical sciences
\cite{mcquarrie,vankampen}. Hence, in the rest of present review
we will not elaborate further on the appropriateness of continuous
representation. This implies that the equations to be discussed
are differential equations. To be more precise, we will postulate
three laws for evolution and the most important law, the second
law, will be expressed in the unique form of stochastic
differential equation. The connection of the present three laws to
well known classical formulations will be discussed. The logical
consistency of the postulated laws will be emphasized, which
results in a flavor of abstractness in the present presentation.

\subsection{What to Be Discussed}

Apart from a concise review of literature, it should be stated
here that the focus of the present article is the structure of
evolutionary dynamics, that is, how many independent dynamical
components should be there, what should be their nature, and which
form the governing dynamical equation should be taken. We will not
discuss the structure of each dynamical component. Making an
analogy to classical physics, what we shall attempt to review and
to establish below is similar to Newton's laws of motion, not of
the structure of mass and force, such as his law of gravity.
Hence, the content of present article corresponds largely to that
when Darwin and Wallace published their theory of evolution by
natural selection \cite{darwin1858}: there are laws of
evolutionary dynamics, but the underlying structures for dynamical
components, such as those related genetics, will not be discussed
but are assumed to exist. This implies that many quantitative and
important problems, such as the speed of evolution
\cite{perlovsky}, cannot be discussed in the present review,
because the knowledge of the dynamical component structure is
needed for such discussions.

\subsection{Organization of Topics}

We organize the rest of the present review style article as
follows. In section II recent efforts in both biological and
physical sciences on the quantitative formulation of evolutionary
dynamics are reviewed. In section III we postulate and discuss the
three laws of evolution. Four dynamical elements will be
introduced and discussed. Fisher's fundamental theorem of natural
selection is reinterpreted as the F-Theorem. In section IV the
connection of the postulated three laws to classical formulations
is discussed. An example of such a connection will be explicitly
demonstrated in section V. The F-Theorem is further discussed in
section VI. Implications of present formulation are discussed in
section VII. Further notes on selected references are presented in
section VIII. Section IX summarizes the results of the present
article.

\section{Efforts on Quantitative Formulation of Evolutionary Dynamics }

In this section we review the previous efforts on the formulation
of evolutionary dynamics. No complete, and even no comprehensive,
coverage of literature is claimed. Nevertheless, we do wish to
convey the message that there have been considerable progress and
that controversies still exist.

\subsection{Ronald Alymer Fisher and Sewall Wright}

Fisher and Wright are among most prominent figures who laid down
the mathematical foundation for the modern quantitative analysis
in population genetics \cite{founders}. Their seminar works were
among those that marked the start of neo-Darwinian synthesis. In
1930, Fisher proposed the fundamental theorem of natural selection
\cite{fisher}, likening it to the second law of thermodynamics in
physics. In 1932, Wright proposed the evolutionary adaptive
landscape in attempting to quantify and to visualize the fitness
in evolutionary dynamics \cite{wright}. The mathematical
developments in population genetics since then have been recently
summarized, for example, by Ewens \cite{ewens} in an effort to
discuss and to contrast the views of Fisher and Wright and to find
a common basis for further development of the field. However, a
considerable amount of controversies still exist around those
central concepts of Fisher and Wright.

It is a consensus that there is no final formulation of Fisher's
fundamental theorem of natural selection. Ewens himself has
changed his interpretation of this theorem between the first and
second editions of his book and considers that the fundamental
theorem of natural selection is still in the developmental stage.
According to Ewens, the original formulation by Fisher was very
obscure, but Ewens holds the view that the theorem is indeed
fundamental. He analyzed various concrete cases to illustrate the
broad implications of this theorem \cite{ewens}. It may be
concluded that in all those examples there is a common thread: the
equivalence between the increase in fitness and the variation in
evolution. Same subject has been also extensively discussed by
Michod \cite{michod} and similar conclusion has been reached.

The adaptive landscape concept of Wright is in a slightly better
situation. Its essence has been elegantly summarized by Ewens
\cite{ewens}, if interpreting the ``the surface of mean fitness" as
the adaptive landscape:

 ``Wright proposed a three-phase process under which evolution
could most easily occur. This view assumes that large populations
are normally split up into semi-isolated subpopulations, or demes,
each of which is comparatively small in size. Within each deme
there exists a genotypic fitness surface, depending on the genetic
constitution at many loci, and in conformity with the `increase in
mean fitness' concept, gene frequencies tend to move so that local
peaks in this surface are approached. The surface of mean fitness
is assumed to be very complex with multiplicity of local maxima,
some higher than others. If a full deterministic behavior obtains
the system simply moves to the nearest selective peak and remains
there. The importance of the comparatively small deme size is that
such strict deterministic behavior does not occur: random drift
can move gene frequencies across a saddle and possibly under the
control of a higher selective peak. Random changes in selective
values can also perform the same function."

Here the important concept of the adaptive landscape implied in
the ``surface of mean fitness" has already been widely used in
population genetics and is particularly illustrative in the
discussion of speciation \cite{ddmt,gavrilets,kisdi,stewart}. Such
metaphoric and visualizing part of Wright's concept has prevailed
in population genetics in particular and biology in large, and has
even permeated into other fields such as statistical physics
\cite{sherrington} and cosmology \cite{smolin}. Connections
between those fields and Darwinian dynamics were already
recognized by Lord Kelvin and by Boltzmann, two of the greatest
physicists in 19th century, during the time of Darwin. It is
interesting to note that ideas inspired by Darwinian dynamics have
also been used in modern quantum physics \cite{zurek}.
Nevertheless, the quantitative part of Wright's concept has
remained an open question in biology. For example, the statement
that no simple explicit formula is known for the stationary
distribution \cite{ewens} may be viewed as such a doubt.

A more explicit negative statement can be found in a recent
textbook by Rice on mathematical foundation of population
genetics. It is based on a perceived absence of potential in limit
cycle dynamics: ``In fact, there is no general potential function
underlying evolution. All that we need to do in order to
demonstrate this is find a case in which, under selection alone,
the allele frequencies in a population do not settle down to a
stable point, but rather continue changing forever. We have
already see an example of this in Figure 1.2. The fact that
selection can result in limit cycles (see Figure 1.2B), in which
the population repeatedly revisits the same states of some
function that increases every generation. Note that this is not a
contradiction of the fundamental theorem, since the
frequency-dependence that drives the fluctuations is part of the
$E(\delta w)$ term in Equation 1.52. Though evolutionary theory is
not built on the idea that any quantity is necessarily maximized,
the idea that there is such a quantity remains one of the most
widely held popular misconceptions about evolution" \cite{rice}.
We will come back to Rice's statement in section V.

We believe that one of key ingredients to properly understand the
works of Wright and Fisher is to re-interpret and to express them
precisely in the light of the modern context of physics and
mathematics, and to show that they are already implicitly
contained in the classical formulations. This will be carried out
in section III and IV.

\subsection{Ecological Evolution }

The evolutionary dynamics from the ecological point of view has
been explored in rather detail. The efforts have been summarized,
for example, in a recent monograph by Turchin \cite{turchin}. This
author takes the view that general laws should exist and should
give rise to useful results relevant to ecology. In particular,
such a general framework should serve a constructive role from the
following three aspects when confronting biological data: defining
what one wants to explain; translating verbal theories into
explicit mathematical models; and using formal statistical methods
to quantify relative ability of rival models to predict data. Such
a view is apparently valid in a broader context in biology
demonstrated in the present article.

In order to provide a consistent mathematical starting point for
ecological dynamics, Turchin has proposed six postulates, inferred
four principles, and stated three dynamical classes. For example,
his six postulates read: (1) The number of organisms in a
population can change only as a result of births, deaths,
immigrations, and emigrations; (2) Population mechanisms are
individual-based; (3) There exists an upper density bound; (4) At
low resource densities the number of resource individuals
encountered and captured by a single consumer is proportional to
resource density----mass action; (5) The amount of energy that an
individual consumer can derive from captured resource, to be used
for growth, maintenance, and reproduction, is a function of the
amount of captured biomass----biomass conversion; (6) No matter
how high the resource density is, an individual consumer can
ingest resource biomass no faster than some upper limit imposed by
its physiology----maximum consumption rate. The principles of
population dynamics inferred by Turchin provides a basis to
classifying dynamics as qualitative kinds of dynamical behaviors,
the dynamical role of various mechanisms, and null hypotheses for
testing consequences of dynamics. Those postulates and principles
do appear to provide a consistent base for analyzing the
ecological data. We refer readers to Turchin's book for a detailed
exposition.

While we share Turchin's positive view on mathematical laws in
biology, particularly about their existence and importance, there
do exist several important differences on how such a general
formulation should be carried out. On the conceptual level, it
appears that Turchin has not differentiated between the laws of
structure of evolutionary dynamics and the laws of the structures
of constituents in the dynamical equations. This may not be
crucial, because it is possible that in ultimate laws the
structure of dynamics and the structure of the dynamical
components may couple to each other, such as in Einstein's theory
of general relativity where the mass and gravity are coupled to
space and time. Nevertheless, we believe that there is a long way
to reach such a state in biology and that it would be better to
start with the situation that those two aspects of laws are
decoupled. Another important difference is that the crucial role
played by stochasticity has not been emphasized in Turchin's
formulation. As a result, the evolutionary potential and Fisher's
fundamental theorem of natural selection are not among Turchin's
explicit topics.

On the technical level, it appears that the number of postulates
in Turchin's formulation is too large and the postulates are too
detailed even for the ecological dynamics. However, none of these
differences consists of an objection to the truth in Turchin's
formulation. They are rather the matter of choosing starting point
from the mathematical point of view, that is, which set of axioms
would be more convenient. The present formulation of population
genetics will start from a different set of axioms. We believe
that the same set of axioms also applies to the formulation of
ecological dynamics. We will come back to Turchin's formulation on
his first and most important principle after the exposition of our
formulation of evolutionary dynamics in section VIII.
%
%
%
%
%

\subsection{Ilya Prigogine }

The evolutionary dynamics may be classified into the realm of
nonequilibrium processes. In this regard, no contemporary scholar
has done more than what Prigogine had done. He had long recognized
that nonequilibrium may be a source of order. He had also
extensively investigated and forcefully generalized the physics'
equivalence of Fisher's fundamental theorem of natural selection,
the fluctuation-dissipation theorem, first systematically
formulated by Onsager in 1931 \cite{onsager}. The importance of
fluctuation has been summarized in his statement that both the
dissipative structure and order are generated through
fluctuations. His work had been condensed into his visionary
theory of dissipative structure. A best place to gain insights
into his theory and the work of his school may be the monograph
written by Nicolis and Prigogine \cite{prigogine}.

The compatibility of physical and biological sciences had been
emphasized by Prigogine, if somewhat biased towards the former. It
is stated \cite{prigogine} that ``far from being outside nature,
biological processes follow from the laws of physics, appropriate
to specific nonlinear interactions and to conditions far from
equilibrium" and that ``thanks to these specific features, the flow
of energy and matter may be used to build and maintain functional
and structural order". Those understandings have now formed the
basis in the study of biological processes. However, even with
such a strong conviction, one of most important concepts in
physics, the potential, has been concluded not applicable in
general \cite{prigogine}. We will come back to this point in section V.
It
should be mentioned that the approach of the Prigogine school is
the best example of the classical approaches discussed below.

\subsection{ Potential in Nonequilibrium Dynamics }

The usefulness of a potential function in a nonequilibrium process
has long been fully recognized in physical sciences. As will be
discussed in the present article, such a quantity can give both
quantitative and qualitative description of behaviors near steady
state or metastable states, such as the distribution function, the
life time of metastabe state, etc. Despite the negative opinion on
its existence reached by many researchers \cite{prigogine,cross},
great efforts have been spent to find such a potential function
and to implement it in applications ever since the work of Onsager
in 1931 \cite{onsager}.

Till 1990's, results of such efforts have been summarized, for
example, in two excellent books by van Kampen \cite{vankampen} and
by Risken \cite{risken}, respectively. Nevertheless, in general
such an effort has been regarded as not very successful
\cite{cross, prigogine}. The situation may be stated as follows.
First, there was no general useful method for the construction of
a potential function. Second, if viewing the log of the steady
state distribution function as potential, it is not clear how this
potential can be related to the dynamic trajectory of the system.
Phrasing in a different way, it is known that the extremal points
of so defined potential function do not necessary coincide with
the fixed points of the drift force. How and why such a mismatch
occurs has not been well understood yet.

The concept of potential in both dynamical and statistical
contexts has been so useful that many researchers simply believe
in its existence. There have been continuous efforts on the
construction of potential function and related topics despite the
doubt \cite{cross}. Nice results have been obtained in several
directions. Tanase-Nicola and Kurchan \cite{kurchan} has
explicitly considered the saddle points of gradient systems by
extending a super-symmetry method. They started from the existence
of potential function to avoid the most difficult problem of the
irreversibility: breaking of the time-reversible symmetry. The
result is a powerful computational method to count the saddle
points and to compute the escape rate.

The study of the mismatch of the fixed point of the drift force
and those of extremals of steady state distribution has been
reviewed by Lindner {\it et al.} \cite{lindner} in the context of
excitable systems widely encountered in physics, engineering, and
biology. Rich phenomena have been observed by computer
simulations. The mismatch has been treated as an numerical
experimental discovery: At this moment there is no mathematical or
theoretical explanation on why and how it should happen.

In another survey the useful and constructive role played by the
noise has been summarized \cite{wio}. It is argued there that the
noise should play an important role in bringing out useful
functions and order, much in the same reasoning as that of Nicolis
and Prigogine \cite{prigogine} but emphasizing on the recent
progress. Again, the above mismatch problem is encountered and the
constructed potential function is typically regarded as
approximation.

From an apparently different direction, there has been an effort
to provide a solid foundation for non-equilibrium processes based
on so-called chaotic hypothesis. Under the chaotic hypothesis an
interesting and important fluctuation theorem has been obtained,
which further suggests the existence of the Boltzmann-Gibbs type
steady state distribution function. Hence, a potential function is
very likely to exist under this situation. The difficulty with
this approach is that extremely few real physical systems have
been shown to satisfy the chaotic hypothesis. It is not clear how
the mathematical construction connects to the physical world. Such
effort has been reviewed by Evans and Searles \cite{evans}.

Because the metastability is such an important phenomenon and
because of the difficulty encountered in the construction of
potential function, effort has been made to go around the
potential function problem but still to be able to compute the
life time of metastable states.  Such endeavor results in the
so-called Machlup-Onsager functional method, summarized by
Freidlin and Wentzell \cite{freidlin}. It has been actively
pursued \cite{maier,beri}. It is an open question that how does
this approach connect to those based on potential function, such
as represented by Kramers' escaping rate formulae \cite{kramers}.

It is clear that the research on the construction of potential in
nonequilibrium processes mirrors the effort on the adaptive landscape
in biology.


\section{Laws of Evolutionary Dynamics}

In this section we postulate and discuss three laws in
evolutionary dynamics. A few of their immediate and important
empirical consequences will be discussed. They form a quantitative
and consistent mathematical framework for evolutionary dynamics.
We also discuss Fisher's fundamental theorem of natural selection
and reformulate it as the F-Theorem. In particular, we will show
that it is indeed an indispensable relationship in the present
formulation.

The rationales to choose the postulation form of presentation are
as follows.

On the positive side, there exist ample empirical evidences that
the domain of population genetics is self-contained. Postulating
the laws would be a direct manner to acknowledge this autonomous
nature of population genetics. This also allows us to introduce
the Wright's adaptive landscape and Fisher's fundamental theorem
of natural selection at the very beginning. Then we can show that
the present formulation is equivalent to the classical
formulations. The ultimate importance and correctness of the
present formulation lies in its ability to connect to empirical
facts. We will point out such connections along with our
presentation.

On the other hand, there exist no rigorous derivations that the
basic mathematical equations in population genetics can be
obtained from more fundamental and microscopic laws in physics and
chemistry. These equations in population genetics, which will be
named classical formulations in the present review, do appeal
reasonable, natural and well supported by empirical facts. We take
the note that there exists no evidence either that they are
incompatible with laws in physics and chemistry. Because we will
not attempt the derivations of the present three laws, we have to
choose a suitable starting point. In addition, it has been
difficult to make the Wright's adaptive landscape and Fisher's
fundamental theorem of natural selection explicit from the
classical formulations, because they have been controversial since
they were proposed seventy years ago as discussed in subsection
II.A and II.D.

The numbering of postulated laws is arbitrary, with an arbitrary
criterion to place the important one in the middle. We start with
the most important and quantitative law, the second law.

\subsection{ Second Law: the Law of Darwin }

The central questions are how to describe the evolutionary
dynamics quantitatively and what are the dynamical elements. To be
specific, let us consider an $n$ component biological system. The
$n$ components may be the species in an evolutionary game
\cite{maynardsmith,turchin}, the traits to describe the speciation
\cite{michod,stewart}, genes in the description
\cite{kisdi,ewens}, or any quantities required to specify the
system. The additional example is that ${\bf q}$ can be the
independent genotypic frequencies in population genetics
\cite{ewens}. The value of $j^{th}$ component is denoted by $q_j$.
The $n$ dimensional vector ${\bf q}^{\tau} = (q_1, q_2, ... ,
q_n)$ is the state variable of the system.  Here the superscript
$\tau$ denotes the transpose. The dynamics of the state variable
is described by its speed $ \dot{\bf q} \equiv d {\bf q}/d t $
moving in the state space.

It is known that the Darwinian evolutionary dynamics is highly
complex \cite{darwin1858,darwin1958,gould,mayr2004,monod}: the
variation exists; there is a selection; and the dynamics is
adaptive. To adequately capture its characteristic, we postulate
that the dynamics of the system in evolution is governed by a
stochastic differential equation, which consists of four dynamical
elements. These are the positive semi-definite symmetric ascendant
matrix $A$, the anti-symmetric transverse matrix $T$, the scalar
function named the Wright evolutionary potential $\psi$, and the
stochastic drive ${\bf \xi}$. The equation reads:
\begin{equation}
  [ A({\bf q}) + T ({\bf q}) ] \dot{\bf q}
     = \nabla \psi ({\bf q}) + {\bf \xi}({\bf q}, t) \; .
\end{equation}
This equation is supplemented by the following relationship
between the stochastic drive and the ascendant matrix:
\begin{equation}
  \langle {\bf \xi}({\bf q}, t)
          {\bf \xi}^{\tau} ({\bf q}, t') \rangle
    = 2 A({\bf q}) \; \epsilon  \; \delta(t-t') \; ,
\end{equation}
and $\langle {\bf \xi}({\bf q}, t) \rangle = 0$. The average $
\langle ... \rangle $ is carried over the dynamics of the
stochastic drive, not over ${\bf q}$, and $\nabla$ is the gradient
operator in the state space. To ensure the independence of the
dynamics of each component, we assume
\begin{equation}
 \det [A({\bf q}) + T({\bf q})] \neq 0 \; .
\end{equation}
Here $\delta(t)$ is the Dirac delta function. In Eq.(2) we have
assumed that the stochastic drive is a Gaussian white noise with
zero mean. Factor 2 is a conventional choice for the present
formulation, and $\epsilon$ is a positive numerical constant,
which for many situations can be set to be unity, $\epsilon = 1$,
without affecting the biological description. The relationship
between the stochastic drive and the ascendant matrix expressed by
Eq.(2) guarantees that the ascendant $ A({\bf q})$ is positive
semi-definite and symmetric.

We leave the more detailed discussion on its connection to
empirical data in section IV when discussing its equivalence to
the classical formulations. In the remainder of this subsection a
few the immediate and important consequences will be discussed.

It is straightforward to verify that the symmetric matrix term is
`ascendant':
\begin{equation}
  \dot{ \bf q}^{\tau} A({\bf q}) \dot{\bf q} \geq 0 \; .
\end{equation}
Its dynamical effect is to increase the Wright evolutionary
potential $\psi({\bf q})$. Here we point out that the graphic
representation of the Wright evolutionary potential corresponds to
the adaptive landscape originally conceived by Wright. Here and
below we avoid the loaded term {\it fitness} whenever possible.
The Wright evolutionary potential defined here has clear meanings
in both dynamics and equilibrium, while the fitness sometimes does
not. The metaphor of adaptive landscape is extended to include all
contributions which may affect the evolution in the considered
process.  The ascendant matrix enables the systems to have the
tendency to seek the largest potential maximum (highest potential
peak). This feature will be explicitly manifested in the
discussion after the first law. The anti-symmetric matrix leads to
no change in potential:
\begin{equation}
  \dot{\bf q}^{\tau} T({\bf q}) \dot{\bf q} = 0 \; ,
\end{equation}
therefore it is conservative. Dynamically it tends to keep the
system along the equal evolutionary potential contour. A
manifestation of the transverse dynamics is the oscillatory
behavior. The effect of the stochastic drive ${\bf \xi}({\bf q},
t)$ on the motion in the adaptive landscape is random: It may
either increase or decrease the evolutionary potential, that is,
it may move up or down in the adaptive landscape. With the above
interpretation, the steady state effect of natural selection is
represented by the gradient of evolutionary potential, $\nabla
\psi({\bf q})$. The clear and graphical discussion of the Wright
evolutionary potential was first explicitly expressed by Wright
\cite{wright} in his concept of adaptive landscape: the potential
peaks corresponding to the Wright's fitness peaks. The
evolutionary potential introduced here is a generalization of
Wright's original concept. The tempo of natural selection is
represented by the ascendant and transverse matrices. Eq.(1)
states that the four dynamical elements, the gradient of Wright
evolutionary potential, the stochastic drive, the ascendant
dynamics, and the transverse dynamics, must be balanced to
generate the evolutionary dynamics.

Eq.(1) is the fundamental equation of evolutionary dynamics
expressed in the unique form of stochastic differential equation:
a form to our knowledge has neither been used in population
genetic nor been generally discussed in mathematically literature.
In accordance with above discussion on stochastic drive and
ascendant matrix, we may call the supplementary equation, Eq.(2),
the stochasticity-ascendancy relation. We will discuss it in the
last subsection in this section in connection with Fisher's
fundamental theorem of natural selection, and suggest to name
Eq.(2) the F-Theorem.

The Wright evolutionary potential $\psi$ introduced in the present
review is in general a highly nonlinear function of state
variables (the populations in ecology and gene frequency in
population genetics). Its role is similar to that of energy in
physical sciences in several aspects. It is in fact opposite in
sign to the typical potential energy used in physical sciences. It
is independent of time here. More precisely, the time variation of
the Wright evolutionary potential is slow comparing to the time
scales of the problem. Extension of present discussion to
explicitly time depending case can be made. If the Wright
evolutionary potential is bounded above, the stationary
probability density to find the system at ${\bf q}$ in state
space, exists, and is expected to be a Boltzmann-Gibbs
distribution:
\begin{equation}
   \rho({\bf q}, t=\infty) = \frac{1}{Z} \exp \left\{
        \frac{ \psi({\bf q}) }{ \epsilon } \right\} \; .
\end{equation}
Here ${Z} = \int \prod_{i=1}^{n} d{q_i} \exp \left\{ {\psi({\bf
q}) } /{ \epsilon } \right\} $, the partition function in physics,
which is the summation over whole state space and serves as the
normalization factor. Eq.(6) will be derived in next section as
Eq.(38). Or it can be directly verified by inserting it into
Eq.(39). For a reader who is familiar with Langevin equation in
physics \cite{goldstein,vankampen}, with $- \psi$ as energy and
$\epsilon$ as temperature Eq.(6) may appear intuitively obvious.
For a reader who is not familiar with the corresponding physics, a
temporary trust on Eq.(6) till Eq.(38) and (39) or (42) is needed.

We should point out that the dynamical aspects of evolution, the
transverse matrix $T$ and the ascendant matrix $A$, do not
explicitly enter into Eq.(6). This is similar to that of the
energy function in classical physics. This suggests that the
Wright evolutionary potential represents the final and ultimate
selection of the evolutionary process. It describes the eventual
landscape to which the system should be adapted. Nevertheless, the
ascendant and transverse matrices, as well as the stochastic
drive, do affect the time to approach to this steady state. From
Eq.(6) we read that the larger the constant $\epsilon$ is, the
wider the equilibrium distribution would be, and more variation
would be, or, the smaller the $\epsilon$ is, the narrower the
distribution. In this sense we may name $\epsilon$ the evolution
hotness. The existence of such a Boltzmann-Gibbs type distribution
suggests a global optimization.

There are a few immediate and important conclusions to be drawn
here. Near a evolutionary potential maximum, corresponding to the
usual Wright evolutionary potential peak, say at ${\bf q}={\bf
q}_{peak}$, we may expand the potential,
\begin{equation}
 \psi({\bf q}) = \psi({\bf q}_{peak}) - ({\bf q} - {\bf
  q}_{peak})^{\tau} U ({\bf q} -{\bf q}_{peak})/2 + O(|{\bf q} -{\bf
  q}_{peak}|^3) \; .
\end{equation}
Here $U$ is a positive definite symmetric matrix as a consequence
at the potential peak. The stationary probability density to find
the system near this peak is typical Gaussian distribution:
\begin{equation}
  \rho({\bf q}, t=\infty) \propto \;  \exp \left\{ - \frac{
   ({\bf q} - {\bf q}_{peak})^{\tau} U ({\bf q} - {\bf q}_{peak}) }
   {2\epsilon} \right\} \; .
\end{equation}
Thus, away from the potential peak, the probability to find the
system will be exponentially small. Such behavior has long been
observed in many biological models
\cite{burger,ewens,felsenstein}.

One may then wonder about how does the system move from one
potential peak to another? This process was first visualized by
Wright \cite{wright}. The relevant mathematical calculation seems
to be first performed by Kramers \cite{kramers}. It had been
applied to biology \cite{barton}, where it was shown that the
stochastic drive must be involved. Hopping from one potential peak
to another must be aided by the stochastic drive. The dominant
factor in the hopping rate $\Gamma$ is usually the difference in
potential between the starting peak (peak1) and the highest point
(saddle point ${\bf q}_{saddle}$) to cross the valley to another
peak \cite{kramers,barton,vankampen}:
\begin{equation}
 \Gamma = P_0( A,T,\epsilon,{\bf q}_{peak1},{\bf q}_{peak2},{\bf q}_{saddle} ) \;
   \exp \left\{ - \frac{\psi({\bf q}_{peak1}) - \psi({\bf q}_{saddle}) }
        {\epsilon }\right\} \; .
\end{equation}
This rate can easily be exponentially small when the exponent
becomes large. The prefactor $P_0$ is usually a smooth function of
the ascendant matrix $A$, the transverse matrix $T$, the numerical
constant $\epsilon$, the positions of starting and ending peaks
and the saddle. The expressions for $P_0$ in several cases have
been explicitly worked out \cite{kramers,vankampen}. Eq.(9) is a
quantitative measure of robustness and stability. Hence it may
explain the usual observation that species are rather stable. Here
species are identified with a particular peak. Nevertheless,
Eq.(9) implies the possibility of the system to move between peaks
when the stochastic drive is finite.

It is already obvious that the shifting balance process visualized
by Wright, as summarized by Ewens \cite{ewens} ({\it c.f.}
subsection II.A), is an apt verbal statement of Eq.(9). We would
like to extend this analogy further. The smaller population size
and a semi-geological isolated group (deme) used by Wright
certainly enhance the hopping rate of the process because of the
enhanced fluctuation. Such process has been mathematically
carefully studied in physics when discussing the transition from a
metastable state to a more stable state: it is energetic very
unfavorable to require the whole system makes a transition at the
same time; rather, it is more feasible to consider such a
transition first occurring locally, in a spatial confined region.
Such a process has been called the nucleation process in physics,
chemistry, and engineering \cite{langer}.

We emphasize the important qualitative feature of Eq.(9). The
hopping rate between peaks is a combined consequence of both the
Wright evolutionary potential $\psi$, the static landscape if we
can neglect its time variation, the stochastic force $\xi$, and
the ascendant matrix $A$. The measure of the strength of
stochastic force, $\epsilon$, is directly appeared in the exponent
in Eq.(9) as the scale for the Wright evolutionary potential.

The second law as expressed by Eq.(1) and (2) captures the major
features of the evolution dynamics first described by Darwin and
Wallace \cite{darwin1858}, extensively exposed by Darwin
\cite{darwin1958}, and further developed by Fisher, Haldane,
Muller, and Wright \cite{founders}, and by many others
\cite{burger,felsenstein}. It expresses the evolutionary process
as a tinkering process \cite{jacob}. In this sense we may also
name the second law the law of Darwin. The necessity and chance
\cite{monod} are represented by the Wright evolutionary potential
and the stochastic drive, respectively.

\subsection{ First Law: the Law of Aristotle }

The first law reflects the case when there is no stochastic drive
in the evolution, i.e., $\epsilon = 0$. This is clearly an
approximation, because variation is always there \cite{kimura},
though it may be regarded to be small on a certain scale.

Allowing the stochastic drive be negligible, Eq.(1) becomes
\begin{equation}
  [ A({\bf q}) + T ({\bf q}) ] \dot{\bf q}
     = \nabla \psi ({\bf q})  \; .
\end{equation}
Because of the ascendant matrix $A$ is non-negative, the system
will approach the nearest attractor determined by its initial
condition, and stay there if already there. Specifically, because
$ \dot{ \bf q}^{\tau} A({\bf q}) \dot{\bf q} \geq 0 $ and $ \dot{
\bf q}^{\tau} T({\bf q}) \dot{\bf q} = 0 $ (eq.(4) and (5)),
Eq.(10) leads to
\begin{equation}
  \dot{ \bf q} \cdot \nabla \psi ({\bf q}) \geq 0 \; .
\end{equation}
This equation implies that the deterministic dynamics cannot
decrease the evolutionary potential: The speed of state variable
$\dot{ \bf q}$ points in the same direction as the gradient of the
Wright evolutionary potential, $\nabla \psi ({\bf q})$. If the
ascendant matrix is positive definite, i.e. $\dot{ \bf q}^{\tau}
A({\bf q}) \dot{\bf q} > 0 $ for any nonzero $\dot{ \bf q}$, the
potential of the system always increases. Hence, the first law
clearly states that the system has the ability to find the local
adaptive landscape peak represented by the Wright evolutionary
potential, determined by the initial condition. However, the
shifting between different evolutionary peaks would become
impossible in this limit, because the shifting probability
vanishes exponentially according to Eq.(9) when $ \epsilon
\rightarrow 0 $. We note that Eq.(11) implies that the Wright
evolutionary potential is a Lyapunov function in dynamical systems
\cite{guckenheimer}.

We have two remarks here. First, from the mathematical theory of
dynamical systems, there are in general three types of attractors
\cite{guckenheimer}: point, (a)periodic, and chaotic (strange).
The point attractors have been well explored in evolutionary
biology since the work of Wright, corresponding to the Wright
evolutionary potential maxima. Other two types of attractors have
also been observed in biology \cite{akin,hastings,murray,may}. The
periodic attractor or the limit cycle in population genetics
\cite{akin,hastings} will be explicitly discussed in section V.
For point attractors, the transverse matrix can be zero. Then the
ascendant route to the potential maximum would follow the most
rapid ascendant path in the landscape defined by the Wright
evolutionary potential. For other two kinds of attractors, the
transverse matrix cannot be zero.

If we further assume the ascendant matrix $A=0$ in Eq.(10), the
evolutionary dynamics will not change the system's evolutionary
potential, hence it is completely conservative. This is precisely
what can be obtained from the Newtonian dynamics \cite{goldstein}:
Energy is conserved in Newtonian dynamics. Here we tentatively
denote the conservative dynamics as the Newtonian dynamics: the
Newtonian dynamics can be casted into the form of
\begin{equation}
 T ({\bf q}) \dot{\bf q} = \nabla \psi ({\bf q}) \; .
\end{equation}
This leads to our second remark: Based on this consideration one
may be inclined to conclude that Newtonian dynamics is a special
case of the Darwinian dynamics as expressed by Eq.(1) and (2), the
case of vanishing ascendant matrix, $A=0$, and zero stochastic
drive, $\xi =0 $. Many biologists may consider this conclusion
natural \cite{dobzhansky,mayr}.

The tendency implied in Eq.(10) to approach a fixed point
(equilibrium) and to remain there has been amply discussed by
Aristotle \cite{aristotle}, well known in both physics and
biology. The periodic motion might have been known in the
classical Greek time, but the concept of strange attractor was
certainly foreign to Aristotle. Nevertheless, to capture this
tendency to an attractor when $\epsilon = 0$,
\begin{equation}
  {\bf q} \rightarrow {\bf q}_{attractor} \; ,  {\ } {\ }
     t \rightarrow \infty \; ,
\end{equation}
and the ability to remain there we may call Eq.(10) the law of
Aristotle. It is evident that in the present setting the first law
is mathematically a special case of the second law. This law does
give us a necessary reference point to define species and other
relevant quantities in a clean manner, if stochasticity could be
ignored.

Two comments are in order to avoid a possible misinterpretation.
As emphasized above, only the meaning of approaching to stable
fixed point has been used in naming Eq.(10) the law of Aristotle.
Our first comment is that the conservative part of dynamics was
not understood by Aristotle. The most clear manifestation of the
conservative dynamics was the inertial law, first formulated by
Newton as his first law of motion in physics, which started the
modern science. The second comment is that, though we assume it to
be zero in Eq.(10), the fluctuation or stochasticity is important.
We postulated it in Eq.(2) that the ascendancy matrix $A$ is
connected to the stochastic drive $\xi$. It was Darwin who first
fully recognized the importance of the fluctuation in biological
evolution, hence the origin and change of species. Those concepts
were not at all appreciated by Aristotle.

\subsection{ Third Law: the Law of Hierarchy }

The third law is a relationship law. It allows us to define the
connection of the focused (mesoscopic, for a convenient
description) level of description to its lower (microscopic) and
higher (macroscopic) ones. For example, there exists at least
three large subfields in biology which mutually influence each
other: ecological dynamics \cite{may,turchin}, population genetics
\cite{ewens,michod}, the focus of the present review, and dynamics
of gene regulatory networks \cite{leibler,zhu2004a,zhu2004b}. Each
subfield may have its own mathematical equations but it is
certainly not isolated from others. For example, all those
dynamics are manifested through the behaviors of each individual
organism of the species. Dynamically, each have its own
characteristic time scale: Ecology in terms of populations of
species may be slow, involving many generations. The continuous
description of population genetics also requires a time span over
many generations. The dynamics of gene regulatory networks is
fast, since it can occur within one generation.

Confining our attention to population genetics, how to make the
connection between the discrete dynamics, the micro level, to the
continuous dynamics, the focused level, is the content of the
third law. It may be stated as follows: The Wright evolutionary
potential $\psi({\bf q})$ has the contribution from lower level in
terms of time average on the time scale of the observation and
description--the level of the present focus, the contribution from
the interaction among various components of the focused level, and
the contribution from the higher level. The stochastic drive
$\xi({\bf q},t)$ is the remainder of all those contributions whose
dynamics is fast on the interested time scale at focused level.
Hence its time average is zero. This stochastic contribution may
be either unknown from microscopic (lower) and mesoscopic and
macroscopic (focused and higher) levels or unnecessary to be
specified in details. Only its probability distribution is needed
and is approximated by a Gaussian distribution in the present
article. The stochastic drive determines the ascendant matrix
$A({\bf q})$, and the transverse matrix $T({\bf q})$ should be
further determined by the dynamics of the system.

The lower, or microscopic, level contribution to the Wright
evolutionary potential $\psi({\bf q})$ and the stochastic drive
$\xi({\bf q},t)$ may allow us to compute the intrinsic adaptive
landscape and the intrinsic source of evolution. However,
historically the computation of this contribution tends to neglect
the horizontal interaction, the interaction among different
populations at the same level of description, which is usually
nonlinear. On the other hand, the same and higher level
contributions may suggest that a control mechanism, such as a
feedback, may be from all three levels. The combination of all
three of them suggests that the evolution is nonlinear,
asymmetric, mutually interactive, stochastic, and may be
controllable.

There is a degree of uncertainty and arbitrariness in both the
assignment of different levels of descriptions and the dichotomy
of deterministic and stochastic terms in Eq.(1). This dilemma has
been amply discussed in both physical \cite{risken,vankampen} and
biological \cite{li,turelli} sciences. This has also been
reflected in the mathematical theory of stochastic process.  Our
way to solve this problem will be proposed as the zero mass limit
in next section in connection to usual dynamics, which is
different from both Ito and Stratonovich approaches and may be
interpreted as an indication of the richness of the hierarchical
structure. The law of the hierarchy requires that the
specification to carrying the analysis of the stochastic
differential equation is needed. Without any such specification,
the Eq.(1) and (2) would be the {\it pre}-equations as discussed
by van Kampen \cite{vankampen}.

The third law is a reflection of the hierarchical structure of the
whole dynamics. The essence of this law is to acknowledge the
existence of multiple time scales: the mesoscopic time scale with
which we observe the system dynamics at the focused level of
description and the micro time scale with which fine structures
and microscopic dynamics come into play. The microscopic time
scale can be regarded as zero in a specific sense.

\subsection{ F-Theorem }

In his classic treatise on genetical foundation of evolution
\cite{fisher}, Fisher stated his fundamental theory of natural
selection (FTNS): {\it The rate of increase in fitness of any
organism at any time is equal to its genetic variance in fitness
at that time}. This statement connects two opposite processes: the
process towards the maximum and the process of variation.

Fisher noticed a profound similarity between FTNS and the second
law of thermodynamics \cite{fisher}:

``Both are properties of
population, or aggregates, true irrespective of the nature of the
units which compose them; both are statistical laws; each requires
the constant increase of a measurable quantity, in the one case
the entropy of a physical system and in the other the fitness,
measured by $m$, of a biological population. As in the physical
world we can conceive of theoretical systems in which dissipative
forces are wholly absent, and in which the entropy consequently
remains constant, so we can conceive, though we need not expect to
find, biological populations in which the genetic variance is
absolutely zero."

His last statement corresponds nicely to the discussion of the
first law in subsection III.B. He likened FTNS to the second law
of thermodynamics, and suggested that it ``should hold the supreme
position among the biological sciences".

Nevertheless, Fisher also noticed five troubling differences
\cite{fisher}, or the exceptions as later called by Gould
\cite{gould}:

``(1) The systems considered in thermodynamics are permanent;
species on the contrary are liable to extinction, although
biological improvement must be expected to occur up to the end of
their existence.

 (2) Fitness, although measured by a uniform method, is
qualitatively different for every different organism, whereas
entropy, like temperature, is taken to have the same meaning for
all physical systems.

 (3) Fitness may be increased or decreased by changes in the
environment, without reacting quantitatively upon that
environment.

 (4) Entropy changes are exceptional in the
physical world in being irreversible, while irreversible
evolutionary changes form no exception among biological phenomena.
Finally,

 (5) entropy changes lead to a progressive disorganization of the
physical world, at least from the human standpoint of utilization
of energy, while evolutionary changes are generally recognized as
producing progressively higher organization in the organic world."

In the present formulation, we notice that the
stochasticity-ascendancy relation, Eq.(2), is similar to a
fluctuation-dissipation theorem (FDT) in physics and chemistry
\cite{onsager,ymr}. FDT may be verbally stated as follows
\cite{kat}: {\it The same processes that drive fluctuations in the
neighborhood of a typical equilibrium or steady state
configuration also drive the configuration back towards a typical
equilibrium when it is displaced from equilibrium}. This is one of
the most important results in nonequilibrium processes. There is a
striking similarity between FTNS and FDT in that the approach to
equilibrium is driven by the fluctuations or variations. Both FTNS
and FDT imply that the return to the evolutionary potential peak
and the fluctuation from the peak are governed by the same
process. Thus, correct interpretation of FTNS should be dynamical
in nature, not thermodynamic. Hence we should replace the fitness
in Fisher's statement by the Wright evolutionary potential. In
addition, we note that the fitness should not correspond to
entropy as Fisher did. Nor should one relate FTNS to the second
law of thermodynamics. Instead, FTNS should correspond to the
stochasticity-ascendancy relation, Eq.(2), an FDT in physics and
chemistry.

With this interpretation of FTNS as FDT, the five exceptions to
FTNS listed by Fisher as quoted above disappear: Since the Wright
evolutionary potential is not entropy, there is no comparison to
the entropy in the physical world, the exception (4) and (5)
immediately become irrelevant. In fact, a nice demonstration on
the difference between biology and physics had been done by
Schr\"{o}dinger \cite{schrodinger}, encoded in his phrase
``negative entropy". The biological systems are open, comparing
with the close systems where the second law of thermodynamics
applies. This important observation has been formulated
quantitatively and in detail by Prigogine in his dissipative
structure \cite{prigogine}.

As indicated by works in population genetics \cite{li,turelli}, a
consideration over many generations is needed to establish the
continuous representation, a manifestation of the third law. The
situation described in exception (1) has been precisely taken care
of. Since the Wright evolutionary potential is not entropy, and
since it is defined for each situations, the exception (2) is
gone. In addition, the Wright evolutionary potential is a relative
concept, there is no meaning to assign an absolute potential
value. It is also a local and individual concept. Instead, entropy
has an absolute value and is global in nature. Because the
evolutionary potential can only be defined through a reference
point, if the environment of a species is changed, even though its
immediate biological consequence is not clear, its evolutionary
potential is changed. Hence, the exception (3) does not apply in
the present formulation.

It should be kept in mind that the attention should not be dwelled
too much on Fisher's exact wording: He formulated FTNS before the
general formulation of FDT in physics and chemistry and his
formulation was close to the modern one: FDT has been indeed
regarded as a dynamical manifestation of the second law of
thermodynamics. In the present reformulation, FTNS represents a
remarkable insight: It immediately connects the adaptation to the
stochastic drive for the motion in the adaptive landscape defined
by the Wright evolutionary potential.

We return to Fisher's original content of FTNS. If the populations
remain fixed and if the state variables are instead the genetic
variables, the variations in Eq.(2) should indeed be the additive
genetic variations. Even in this situation we should emphasize
that the Eq.(2) is independent of the Wright evolutionary
potential. This independence is the indication of its fundamental
importance. For example, if there is no selection by fitness, that
is, if the Wright evolutionary potential is a constant, the system
would diffuse over the whole state space in search of the largest
and simple connected volume in hyperspace according to FTNS, such
as in the case of holey adaptive landscape model for speciation
\cite{gavrilets}.  This suggests that the {\it the rate of
increase} used in Fisher's statement is not precise, which may
only be true near a potential maximum. In general, it is the
diffusion matrix which associates with the stochasticity in the
conventional formulation to be discussed in next section. We will
make this point more concrete in section VI.

As implied in the first law, the ascendancy of the system is
described by the ascendant matrix $A$, which in turn is completely
determined by the stochastic drive $\xi$ according to the
stochasticity-ascendancy relation, Eq.(2). The discussion
following Eq.(9) indicates that the ability of the system to find
a better evolutionary potential maximum, not only the nearest
local maximum, or, to reach the global equilibrium, is guaranteed
by the stochastic drive. This suggests that Eq.(2) is a statement
on the unification of the two completely opposite tendencies:
adaptation and optimization.

It is interesting to point out that even this unification of
ascendancy and stochastic effect had not been completely brought
out by Fisher. Without the stochastic drive in the dynamics
equation, we are effectively dealing with Eq.(10), the first law
in the present formulation. Therefore, the evolution would
eventually lead to the highest local peak, and would stay there.
Mathematically, this implies that eventually  $\dot{\bf q}(t
\rightarrow \infty) = 0$, a fact which is in clear contradiction
to observations. Fisher apparently recognized this difficulty. He
did it away with the continuous changing of his fitness by
`deterioration of the environment' \cite{fisher}, which does not
completely agree with observations either, though it is more
difficult to be tested by observations. In the light of the
present formulation, the including of stochastic drive into the
dynamics equation makes the introduction of `deterioration of the
environment' completely unnecessary. There is a dynamical balance
right at the evolutionary potential peak as described by Eq.(1)
and (2).

Ever since Fisher's proposal of the fundamental theorem of natural
selection, misrepresentation and misunderstanding have been
associated with this insightful statement, as discussed by Crow
\cite{crow} and  Grafen \cite{grafen}. Though Fisher might have
confined his discussion within genetic variations, the comparison
of Fisher's statement to Eq.(2)  displays a remarkable similarity.
We emphasize again that on the left hand side of Eq.(2) is the
measure of the variations, which should include all variations
affecting the evolutionary dynamics, including the additive
genetic variation, and on the right hand side may be interpreted
as the rate to the peak of the Wright evolutionary potential, with
a proper choice of units. Hence, we may also call Eq.(2) the
fundamental theorem of evolution, as a tribute to Fisher's great
insight, or the Fisher's theorem. Due to its similarity to the
fluctuation-dissipation theorem in physics, we may simply name
Eq.(2) the F-Theorem for short, to incorporate all related
meanings. Its importance is evident: It is an integral part of the
second law of evolution, the Darwinian dynamics. It enables the
system to find the global evolutionary potential maximum. It
should be pointed out again that the F-Theorem in the present
formulation is independent of the Wright evolutionary potential:
It exists even if the Wright evolutionary potential is a constant.
By generalization the variation beyond genetic ones, this
fundamental theorem can be also used in other parts of biology,
for example, developmental biology and ecology.

A note on the semantics is in order. We have named Eq.(2) the
F-Theorem in the present subsection, in connection to fundamental
theorem of natural section and to fluctuation-dissipation theorem.
In strict mathematical sense this is inappropriate: Eq.(2) has
been introduced in the present article as a part of a postulate.
It is a part of an axiom, not a derived result in the present
formulation. Nevertheless, just like in the case of the
fluctuation-dissipation theorem we believe that the F-Theorem may
be derivable from a more fundamental description and the word
``theorem" is used here in this sense. For the F-Theorem in the
context of population genetics we nevertheless are not aware of
any such derivation. For the derivation of fluctuation-dissipation
theorem in physics and chemistry, we suggests the interesting
readers to Ref.[\cite{gbg}] and [\cite{evans}] for a taste of the
difficulty, where its mathematically rigorous justification is an
ongoing research program.

\section{ Classical Formulations }

The purpose of this section is to demonstrate that the three laws
proposed in section II are in fact deeply rooted in the classical
formulations of evolutionary dynamics. It will reveal,
particularly in subsection IV.A, that a conservative dynamics
represented by the transverse $T$ is needed to unify ideas from
Wright and Fisher. This conservative dynamics has been overlooked
in the study of evolutionary dynamics. The demonstration of the
equivalence below also provides a procedure, not necessarily easy
in mathematics and in computation, to compute the Wright
evolutionary potential, the ascendant and transverse matrices from
classical formulations.

It is known that there are several standpoints to study the
dynamics. One focuses on the trajectory of state variables as
function of time, Newton's three laws on dynamics are such
example. Another focuses on the distribution of state variable,
such as the Liouville formulation of classical dynamics in physics
\cite{goldstein}. Both are equivalent and complementary to each
other. There is a third standpoint, the discrete description of
dynamics, which can be connected to either of above standpoints
and will not be pursued in the present article. For such a
connection we refer readers to the commentated collected papers in
population genetics \cite{li} and an insightful paper in physical
sciences \cite{mcquarrie}. The previous approach from either the
system's trajectory in state space or the evolution of probability
distribution function point of view will be tentatively named the
classical formulation.

\subsection{ Standard Stochastic Differential Equation }

Now we make the connection between the dynamics described by
Eq.(1) and (2) and the dynamical equations typically encountered
in evolution. We start with the standard stochastic differential
equation widely used in mathematical literature
\cite{turelli,vankampen}:
\begin{equation}
  \dot{\bf q} = {\bf f}({\bf q}) +
                    {\bf \zeta}( {\bf q}, t) \; .
\end{equation}
Here ${\bf f}({\bf q})$ is the deterministic nonlinear drive of
the system, which includes effects from both other components and
itself. The stochastic drive is ${\bf \zeta}( {\bf q}, t)$, which
differs from $\xi$ in Eq.(1) in appearance but both are governed
by the same dynamics. For simplicity we will assume that ${\bf f}$
is a smooth function whenever needed. Without loss of generality
the stochastic drive in Eq.(14) is assumed to be Gaussian and
white with the variance,
\begin{equation}
 \langle {\bf \zeta}( {\bf q}, t)
         {\bf \zeta}^{\tau}({\bf q}, t') \rangle
     = 2 D( {\bf q}) \; \epsilon \; \delta (t-t') ,
\end{equation}
and zero mean, $\langle {\bf \zeta}({\bf q}, t)\rangle = 0$.
Eq.(9) is consistent with Eq.(2), Gaussian white noise and same
dynamical origin. Again here $\langle ... \rangle$ indicates the
average with respect to the dynamics of the stochastic drive $\zeta$,
not over the state variable ${\bf q}$.
According to the biology convention the positive semi-definite
symmetric matrix $D=\{D_{ij}\}$, with $i,j=1,2, ..., n$, is the
diffusion matrix.

Both the divergence and the skew matrix of the nonlinear drive
${\bf f}$ are in general non-zero:
\begin{eqnarray}
 \nabla \cdot {\bf f} & \neq & 0  \; , \\
 \nabla \times {\bf f}& \neq & 0  \; .
\end{eqnarray}
Here the divergence is explicitly $\nabla \cdot {\bf f} =
\sum_{j=1}^{n} \partial f_j / \partial q_j = tr(S)$, and the skew
matrix $\nabla \times {\bf f}$ is twice the anti-symmetric part of
the selection matrix $S$: $( \nabla \times {\bf f} )_{ij} = S_{ji}
- S_{ij} $, with
\begin{equation}
 S_{ij} = \frac{ \partial f_i }{\partial q_j } \ , \; i,j =1,2, ..., n \; .
\end{equation}
Since the divergence is non-zero the state space volume is not
conserved, leading to ascendancy. The skew matrix is again
non-zero, or putting it differently the selection matrix $S$ is
asymmetric. This implies the existence of the transverse matrix
$T$.

A note on notation is in order: In the classical formulations
\cite{ewens} the diagonal part of $D$ is called variance, also
called the diffusion coefficients, and the off-diagonal part of
$D$ is called covariance. Mathematically, a coordinate
transformation exists that diagonalizes the matrix $D$, hence
turns $D$ into diffusion coefficients in the new coordinate
representation. To simultaneously maintain the general expression
for D and to keep the meaning of diffusion, we here name the whole
matrix $D$ as the diffusion matrix. The diffusion matrix $D$ is in
general a function of state variable and time.

Now, we give an explicit construction which demonstrates the
connection between Eqs.(1,2) and Eqs.(14,15). Assume that both
Eq.(1) and (14) describe the same dynamics. The speed $\dot{\bf q}
$ is then the same in both equations. The connection from Eq.(1)
to (14) is straightforward: Multiplying both sides of Eq.(1) by $
[A({\bf q})  + T({\bf q}) ]^{-1} $ leads to Eq.(14). Converting
Eq.(14) into (1) is mathematically more involved.

Using Eq.(14) to eliminate the speed $\dot{\bf q}$ in state space
in Eq.(1), we have
\[
   [A({\bf q}) + T({\bf q})] [ {\bf f}({\bf q})
     + {\bf \zeta}({\bf q}, t) ]
   =  \nabla \psi({\bf q}) + {\bf \xi}({\bf q},t) \; .
\]
The dynamics of stochastic drive is different from that of the
state variable. This is expressed in Eq.(2) or (15) in that the
average is not on state variable. This is also in consistent with
the third law: the stochastic dynamics is from a different level.
Thus, the stochastic drive and the state variable can be regarded
as independent variables. This suggests following two equations
\begin{equation}
  [A({\bf q}) + T({\bf q})] {\bf f}({\bf q})
   =  \nabla \psi({\bf q}) \; ,
\end{equation}
and
\begin{equation}
   [A({\bf q}) + T({\bf q})] {\bf \zeta}({\bf q}, t)
       = \xi({\bf q}, t) \; .
\end{equation}
Those two equations suggest a rotation in state space.

Multiplying Eq.(20) by its own transpose on each side and
averaging over the stochastic drive, we have
\begin{equation}
 [A({\bf q}) + T({\bf q}) ] D({\bf q})
 [A({\bf q}) - T({\bf q}) ] = A({\bf q}) \; .
\end{equation}
In obtaining Eq.(21) we have also used Eq.(2) and (15). Eq.(21)
suggests a duality between the standard stochastic differential
equations and Eq.(1): A large ascendant matrix $A$ implies a small
diffusion matrix $D$ when the transverse matrix $T$ is small. In
the opposite limit of large transverse matrix $T$, the diffusion
matrix $D$ is proportional to the ascendant matrix $A$. It is a
generalization of the Einstein relation \cite{einstein,vankampen}
to nonzero transverse matrix.

Next we introduce an auxiliary matrix function $G$ for a better
connection to the classical formulation, defined as
\begin{equation}
   G({\bf q}) = [A({\bf q}) + T({\bf q}) ]^{-1} \; .
\end{equation}
Here `${-1}$' indicates the matrix inversion. Such an auxiliary
function bring out a close relationship between the presentation
formulation and those of classical ones. In particular, the role
of variance in the classical formulations, the diffusion matrix
$D$, has a straightforward meaning in $G$, exemplified by Eq.(27)
below.

Using the property of the Wright evolutionary potential $\psi$:
\begin{equation}
 \nabla \times \nabla \psi = 0
\end{equation}
[$(\nabla \times \nabla)_{ij} = \nabla_i \nabla_j -\nabla_j
\nabla_i = \partial^2/\partial {q_i} \partial {q_j} -
\partial^2/\partial {q_j} \partial {q_i} \; , i,j = 1,2, ... , n $ ],
Eq.(19) leads to
\begin{equation}
  \nabla \times [ G^{-1} {\bf f}({\bf q}) ] = 0 \; ,
\end{equation}
which gives $n(n-1)/2$ conditions.

Note that
\begin{eqnarray}
  A({\bf q}) & = & \frac{1}{2} [(A({\bf q}) + T({\bf q}))
       + (A({\bf q}) + T({\bf q}))^{\tau}]  \nonumber \\
   & = & \frac{1}{2} [ G^{-1} + (G^{\tau})^{-1} ] \; ,
\end{eqnarray}
and that the generalized Einstein relation, Eq.(21), can be
rewritten as
\begin{equation}
 G^{-1} D (G^{\tau})^{-1}
  = \frac{1}{2} [ G^{-1} + (G^{\tau})^{-1} ] \; ,
\end{equation}
we have thus the following equation
\begin{equation}
   G + G^{\tau} = 2 D \; ,
\end{equation}
which readily determines the symmetric part of the auxiliary
matrix $G$, another $n(n+1)/2$ conditions.  Eq.(24) and (27) form
a complete set of equations: The total number of equations
embedded in them is $n^2$, precisely the number to determine the
auxiliary matrix $G$. Hence, with an appropriate boundary
condition, $G$ can be found, so will be $A$, $T$, and $\psi$. A
gradient expansion scheme to solve Eqs.(24,27) is presented below,
which emphasizes the role played by fixed points: It is exact near
fixed points.

The auxiliary function $G$ may be formally solved as an iteration
in gradient expansion:
\begin{equation}
  G = D + Q \; ,
\end{equation}
with
\begin{eqnarray}
 Q           & = & \lim_{j \rightarrow \infty} \Delta G_j  \; , \\
 \Delta G_j  & = & \sum_{l=1}^{\infty} (-1)^l [ (S^{\tau})^l
   \tilde{D}_j S^{-l} + (S^{\tau})^{-l} \tilde{D}_j S^l ] \; ,  \\
 \tilde{D}_0 & = & D S - S^{\tau}D                    \; , \\
 \tilde{D}_{j \geq 1}
             & = & ( D + \Delta G_{j-1} )
     \left\{ [\nabla \times (D^{-1} + \Delta G_{j-1}^{-1} ) ]
     {\bf f} \right\} ( D - \Delta G_{j-1} ) \; .
\end{eqnarray}
At each step of solving for $\Delta G_j$ only linear algebraic
equation is involved. One can verify that the matrix $Q$ is
anti-symmetric. For a simple case a formal solution of such
algebraic equation was given in \cite{ao2002}, and an explicit
procedure was found for generic cases in \cite{kat}. Eqs.(29-32)
are the result of a local approximation scheme: If the selection
matrix $S$ and the diffusion matrix $D$ are constant in space, the
exact solution only contains the lowest order contribution in the
gradient expansion: $Q= \Delta G_j = \Delta G_0 $. We regard
Eqs.(29-32) as the biological solution to Eq.(24) and (27),
because it preserves all the fixed points of deterministic drive
${\bf f}$. The connection from Eq.(14) to (1) is therefore
uniquely determined:
\begin{equation}
  \left\{ \begin{array}{lll}
  \psi({\bf q}) & = & \int_C d{\bf q}' \cdot
          [ G^{-1}({\bf q}') {\bf f}({\bf q}') ] \\
  A({\bf q}) & = & [G^{-1}({\bf q})+(G^{\tau} )^{-1}({\bf q})]/2 \\
  T({\bf q}) & = & [G^{-1}({\bf q})-(G^{\tau} )^{-1}({\bf q})]/2
  \end{array}
     \right. \; .
\end{equation}
Here the sufficient condition $\det(A+T)\neq 0$ is used, and the
end and initial points of the integration contour $C$ are $\bf q$
and ${\bf q}_0$ respectively.

We point out that in the limit of vanishing strength of stochastic
drive, i.e., $\epsilon = 0$ in Eq.(2) and (15), the above
connection remains unchanged.

\subsection{ Fokker-Planck Equation }

In experimental studies in biology, we are often interested in the
distribution of the state variable versus time rather than
focusing on the individual trajectory of the system. This is
especially true in population genetics where one almost
exclusively deals with distribution. This implies that either
there is an ensemble of identical systems or repetitive
experiments are carried out. To describe this situation, we need a
dynamical equation for the distribution function in the phase
space. This goal can be accomplished by the so-called
Fokker-Planck equation, or the diffusion equation, or the
Kolmogorov equation \cite{li,burger,ewens,risken,vankampen}.

In this subsection, a new procedure is presented to find the
equation of the distribution function. It is motivated from a
theoretical physics point of view. This procedure will establish
that the Wright evolutionary potential $\psi$ in Eq.(1) indeed
plays the role of potential energy in physics in the manner
envisioned by Wright: the adaptive landscape is quantified by the
evolutionary potential and the steady state distribution is given
by Eq.(6). The system indeed tends to stay at the peaks of
evolutionary potential as suggested by the distribution function.
Our starting point will be the second law, Eq.(1), not the
standard stochastic differential equation, Eq.(14), or a master
equation from which most previous derivations in population
genetics started \cite{ewens}.

The existence of both the deterministic and the stochastic drives
in Eq.(1) suggests that there are two well separated time scales
in the system: the microscopic or fine time scale to describe the
stochastic drive and the macroscopic or course time scale to
describe the system motion. The former time scale is much smaller
than the latter. This separation of time scales further suggests
that the macroscopic motion of the system has ``inertial": it
cannot response instantaneously to the microscopic motion. To
capture this feature, we introduce a small constant inertial
``mass" $m$ (no confusion with the $m$ used by Fisher) and a
kinetic momentum vector ${\bf p}$ for the system. Our state space
is then enlarged: it is now a $2n$-dimensional space. The
dynamical equation for the system takes the form:
\begin{equation}
   \dot{\bf q} = \frac{ {\bf p} }{m} \; ,
\end{equation}
which defines the kinetic momentum, and
\begin{equation}
    \dot{\bf p} = - [A({\bf q}) + T({\bf q}) ] \frac{\bf p}{m}
                         + \nabla \psi({\bf q})
                         + {\bf \xi}({\bf q}, t) \; ,
\end{equation}
which is the extension of Eq.(1): In the limit $m \rightarrow 0$
we recover Eq.(1) ({\it c.f.} below). We note that the ascendant
matrix $A$ and the stochastic drive are independent of the kinetic
momentum ${\bf p}$. The Fokker-Planck equation in this enlarged
state space can be immediately obtained \cite{vankampen}:
\begin{equation}
 \left\{ \partial_t + \frac{\bf p}{m} \cdot \nabla_{\bf q}
           + \overline{\bf f} \cdot \nabla_{\bf p}
  - \nabla_{\bf p}^{\tau} A \left[\frac{\bf p}{m} + \nabla_{\bf p}
\right]
\right\}
   \rho({\bf q}, {\bf p}, t) = 0 \; .
\end{equation}
Here
\begin{equation}
  \overline{\bf f} = \frac{ {\bf p}^{\tau} T }{m} +
    \nabla_{\bf q} \psi \; ,
\end{equation}
and $t$, ${\bf q}$, and ${\bf p}$ are independent variables. The
subscripts of $\partial$ and $\nabla$ indicate the differentiation
with respect to the indicated. The stationary distribution can be
found and easily verified, when the Wright evolutionary potential
is bounded above and normalizable, as \cite{vankampen}
\begin{equation}
 \rho({\bf q},{\bf p},t=\infty) = \frac{1}{\cal Z} \exp\left\{-\frac{
   {{\bf p}^2 }/{2m} - \psi({\bf q}) }{\epsilon } \right\} \; ,
\end{equation}
with ${\cal Z} = \int \prod_{i=1}^{n} d{ q_i} \prod_{i=1}^{n} d{
 p_i} \exp\{ - [ {\bf p}^2 /2m - \psi({\bf q}) ]/\epsilon \}$
the partition function in the extended state space. There is an
explicit separation of state variable and its kinetic momentum in
Eq.(38). The elimination of the momentum in the small mass limit
will not affect this distribution. Hence, Eq.(38) confirms that
the expected Boltzmann-Gibbs distribution, Eq.(6) from the Eq.(1)
and (2), is the right distribution function.

We proceed to outline the procedure to find the Fokker-Planck
equation corresponding to Eq.(1) and (2) without the kinetic
momentum ${\bf p}$. We first illustrate how to recover Eq.(1) from
Eq.(34) and (35). In the limit of $m \rightarrow 0$, the fast
dynamics of kinetic momentum ${\bf p}$ always follows the motion
of the state variable ${\bf q}$ whose dynamics is slow. Hence we
may set $\dot{\bf p} = 0 $ in Eq.(35) and replace the kinetic
momentum using Eq.(34), which is then Eq.(1) after moving the
speed to the left-side of equation. For the Fokker-Planck
equation, the explicit separation of the kinetic momentum and
state variable in the stationary distribution gives the guidance
on the procedure: The resulting Fokker-Planck equation must be
able to reproduce this feature. The Fokker-Planck equation is then
found as
\begin{equation}
  \partial_t \rho({\bf q},t)
   = \nabla^{\tau} [ - {\bf f}({\bf q}) - \Delta{\bf f}({\bf q})
                     + D({\bf q}) \nabla ] \rho({\bf q}, t) \; ,
\end{equation}
with $\Delta {\bf f}$ the solution of the equation
\begin{equation}
  \nabla \cdot \Delta {\bf f} + \Delta {\bf f}\cdot
\nabla \psi
  - \nabla \cdot [ G T G^{\tau} \nabla \psi ] = 0 \; .
\end{equation}
%
%
If the probability current density is defined as
\begin{equation}
 {\bf j}({\bf q},t) \equiv ( {\bf f} + \Delta{\bf f}
  - D \nabla ) \rho({\bf q},t) \; ,
\end{equation}
the Fokker-Planck equation is a statement of the probability
continuity:
\begin{equation}
  \partial_t \rho({\bf q},t) + \nabla \cdot {\bf j}({\bf q},t) = 0 \; .
\end{equation}
In population genetics equations in the form of Eq.(39) or (42)
have been routinely and successively employed
\cite{burger,ewens,felsenstein,li}. The stationary state
corresponds to the condition
 $ \nabla \cdot {\bf j}({\bf q}, t=\infty) = 0 $.
One may verify that the stationary distribution $\rho({\bf q},
t=\infty)$ in Eq.(6) is indeed the time independent solution of
the Fokker-Planck equation: The stationary probability current
\begin{equation}
 {\bf j}({\bf q}, t=\infty) = (G T G^{\tau} + \Delta {\bf f})
  \nabla \psi({\bf q}) \; \rho({\bf q},t=\infty) \; ,
\end{equation}
and $\nabla \cdot {\bf j}({\bf q},t=\infty) = 0 $.

The connection between the standard stochastic differential
equation and Fokker-Planck equation has been under intensive study
by biologists, physicists, chemists, mathematicians, and others
over last 70 years \cite{li,burger,felsenstein,vankampen}.
However, there exists an ambiguity for the generic nonlinear
situation \cite{vankampen,turelli}. We attribute this ambiguity to
the asymptotic nature of the connection in which a procedure must
be explicitly defined: Different procedures will in general lead
to different results. Biologically, it is a statement on how the
dichotomy of deterministic and stochastic drives is done, a
genuine indication of the hierarchical nature of the dynamics.
What has been demonstrated in this subsection is one way of
carrying out this procedure:
\begin{equation}
  m \rightarrow 0 \; .
\end{equation}

Mathematically, we have now demonstrated the equivalency between
Eq.(1) and (2) and the classical formulations in terms Eq.(14) and
(15) or Eq.(39): there indeed exist evolutionary potential, the
ascendant matrix, and the transverse matrix. The hierarchical
structure is represented by the zero mass limit $m \rightarrow 0$.
The four dynamical components are all mixed up in the classical
formulation of Eq.(14) and (15). For the classical formulation
represented by eq.(39) or (42), not only four dynamical components
are mixed, the zero mass limit is also needed. Hence Eq.(39) or
(42) effectively encodes all the three laws postulated in section
III. From Eq.(1) and (2) to Eq.(14) and (15) it is straight
forward: involving only algebraic manipulation. The converse is in
general not easy: the task to explicitly construct the
evolutionary potential from either Eq.(14) and (15) to Eq.(1) and
(2) involving solving first order partial differential equation
with appropriate boundary conditions. This difficulty has been
encountered in previous attempts and is the reason that there are
only limited successes \cite{cross,prigogine}. In the next
section, section V, a nontrivial example will be explicitly
discussed.

There is a question of the meanings of kinetic momentum $p$ and
the mass $m$. Their meanings in classical physics are well defined
and will not be elaborated here. We refer readers to
Ref.[\cite{goldstein}] for an authoritative discussion. However,
it is not at all clear what are their precise meanings in
population genetics. They have been used here as mathematical
devices to make the connection between the present formulation of
Eq.(1) and (2) to the classical formulations and to show that the
evolutionary potential and other quantities do exist. They are
also used as a way to realize the hierarchical structure in
population genetics. In the end they do not explicitly appear in
dynamical equations. Whether or not they have deeper meanings in
population genetics in particular and in biology in general
remains to be explored ({\it c.f.} subsection VII.C.).

\subsection{ Detailed Balance Condition }

There is an important class of evolution dynamics in which the
anti-symmetric matrix $Q = 0$. Under this condition, the
transverse matrix $T=0$, and  $\Delta {\bf f} = 0$. The
Fokker-Planck equation becomes
\begin{equation}
  \partial_t \rho({\bf q},t)
   = \nabla^{\tau} [ - {\bf f}({\bf q}))
                     + D({\bf q}) \nabla ] \rho({\bf q}, t) \; ,
\end{equation}
and the stationary probability current is everywhere zero in state
phase:
\begin{equation}
   {\bf j}({\bf q}, t=\infty) = 0 \; .
\end{equation}
In this situation one finds that
\begin{eqnarray}
 \nabla \psi({\bf q}) & = & D^{-1}({\bf q}) {\bf f}({\bf q}) \; , \\
  A  & =  & D^{-1} \; .
\end{eqnarray}
The Wright evolutionary potential and the connection between
Eq.(1) and the standard stochastic differential equation can be
directly read out from equations. This is the well-known symmetric
dynamics in biological \cite{li,burger,felsenstein} and physical
\cite{vankampen} sciences. A condition to generate this kind of
equilibrium state in biology was first noticed, in retrospective,
by Hardy \cite{hardy} and Weinberg \cite{weinberg}. This zero
probability current condition is the usual detailed balance
condition.

\subsection{More on the Third Law}

We may further illustrate the third law. In reaching Eq.(39) we
have taken a specific way to classify the stochastic and
deterministic drives, the zero ``mass" limit, $m \rightarrow 0$.
We mentioned in subsection III.C that there are other ways to do
this classification. Two of them are particularly well known in
physical and biological sciences, Ito and Stratonovich. Starting
from Eq.(14) an d(15), the Fokker-Planck equation according to
Ito's prescription is \cite{vankampen}
\begin{equation}
  \partial_t \rho({\bf q},t)
   = \nabla^{\tau} [ - {\bf f}({\bf q}) - \Delta{\bf f}_{Ito}({\bf q})
                     + D({\bf q}) \nabla ] \rho({\bf q}, t) \; ,
  {\ } {\ } {\ } (Ito)
\end{equation}
and the $i$-th component of correction $\Delta {\bf f}_{Ito}({\bf
q})$
\begin{equation}
 \Delta{f}_{Ito,i}({\bf q})
  = - \sum_{j=1}^{n} \nabla_j D_{ij}({\bf q})     \; .
\end{equation}
Starting from Eq.(14) and (15), the Fokker-Planck equation
according to Stratonovich's prescription is \cite{vankampen}
\begin{equation}
  \partial_t \rho({\bf q},t)
   = \nabla^{\tau} [ - {\bf f}({\bf q}) - \Delta{\bf f}_{Str}({\bf q})
                     + D({\bf q}) \nabla ] \rho({\bf q}, t) \; ,
   {\ } {\ } {\ } (Str)
 \end{equation}
and the $i$-th component of correction $\Delta{\bf f}_{Str}({\bf
q})$
\begin{equation}
 \Delta{f}_{Str,i}({\bf q})
  = - \sum_{j,l=1}^{n} d_{il}({\bf q}) \nabla_j d_{jl}({\bf q})     \; .
\end{equation}
Here we have rewritten the standard stochastic differential
equation, Eq.(14), in the form
\begin{equation}
  \dot{\bf q} = {\bf f}({\bf q}) +
                    d({\bf q}){\bf \hat{\zeta}}(t) \; ,
\end{equation}
with $d({\bf q})$ is a smooth $n \times n$ matrix function of
${\bf q}$ to remove the state variable dependence from the noise
${\bf \zeta}({\bf q},t)$,
\begin{eqnarray}
 D({\bf q}) & = & d({\bf q}) d^{\tau}({\bf q}) \; , \\
 \langle\hat{\zeta}_i(t) \hat{\zeta}_j(t') \rangle
            & = & 2 \delta_{ij} \delta(t-t') \; ,
\end{eqnarray}
and $\langle \hat{\zeta} \rangle = 0$,

Eq.(39), (49), and (51) are all different from each other. The
major difference lies in the deterministic driving force ${\bf
f}$: after a proper re-definition of driving force they all can be
rewritten in the form of Eq.(39) with different $\Delta{\bf
f}({\bf q})$ and different ${\bf f}$ to correspond back to Eq.(1)
and (2). Such a mathematical reconstruction is similar to the
Legendre transformation in thermodynamics and will be explored
elsewhere. Those equations, Eq.(39), (49), and (51), suggest that
in reaching Eq.(39), the third law postulated in section III is
necessary.

It is evident that all four dynamical elements are implicitly
included in Eq.(39). Starting from Eq.(39), in order to
reconstruct the four dynamical elements, we may need to find the
corresponding Eq.(14) and (15) first. Then using Eq.(24) and (27)
for a further reconstruction to obtain Eq.(1) and (2). This
suggests that the Wright's evolutionary potential always exists,
contrast to the negative claim in literature \cite{rice}. Examples
of such reconstruction can be found in Ref.[\cite{kat}] for fixed
points and in next section for a limit cycle dynamics.

\section{Evolutionary Potential and Limit Cycle Dynamics}

In this section we discuss an intriguing case of co-existence of
limit cycle dynamics and the evolutionary potential. The limit
cycle dynamics exists in biological fields, ecology
\cite{murray,turchin}, population genetics
\cite{akin,hastings,rice}, or gene regulatory networks
\cite{leibler,prigogine}. The existence of evolutionary potential
in population genetics has been questioned because of a perceived
absence of potential (Lyaponuv function) in limit cycle
\cite{rice}. We will explicitly show the existence of evolutionary
potential in a limit cycle and demonstrate its generic
implication. It is an illustration on the connection between the
proposed laws discussed in section III and the classical
formulations in section IV. The example also serves an indication
for the practical difficulty in the construction of the
evolutionary potential out of the classical formulations.

\subsection{Limit Cycle in Physics}

The goal of this subsection is twofold: to provide a concrete
anchoring point where all the quantities are physically
conceivable and to obtain an explicit example of potential with
expected properties in the presence of limit cycle.

In physics, the general dynamical equation for a {\it massless}
particle in two dimensional state space may be expressed, with
both deterministic and stochastic forces \cite{goldstein}:
\[
  [ A({\bf q}) + T ({\bf q}) ] \dot{\bf q}
          =    \nabla \psi ({\bf q}) + {\bf \xi}({\bf q}, t) \; ,
\]
identical to Eq.(1), and supplemented by the relationship on the
stochastic force:
\[
  \langle {\bf \xi}({\bf q}, t) {\bf \xi}^{\tau} ({\bf q}, t') \rangle
         = 2 A({\bf q}) \; \epsilon  \; \delta( t-t' ) \; ,
\]
and $\langle {\bf \xi}({\bf q}, t) \rangle = 0$, identical to
Eq.(2). Here ${\bf q}^{\tau} = (q_1, q_2)$ with $q_1, q_2$ the two
Cartesian coordinates of the state space, which may be perceived
as the position space of the massless particle. The transpose is
denoted by the superscript $\tau$, and $\dot{\bf q} = d {\bf q}/dt
$.

The scalar function $\psi$ corresponds to the usual potential
energy function. Its graphical representation in the state space
is a landscape. The antisymmetric matrix $T$ represents the
dynamics which conserves the potential, corresponding to the
Lorentz force in physics, determined by the magnetic field. The
matrix $A$ represents the dynamics which increases the potential
(Please note that the evolutionary potential has an opposite sign
as that in physics.), the dissipation in physics. This matrix may
be called the friction matrix as used in physics. The simplest
realization of the friction matrix is a constant matrix
\begin{equation}
 A = \eta \left(
    \begin{array}{ll}
               1 & 0  \\
               0 & 1
             \end{array} \right)
\end{equation}
with the friction coefficient $\eta$. The simplest realization of
the antisymmetric matrix is
\begin{equation}
 T  = b \left(
  \begin{array}{ll}
                  0 & -1  \\
                  1 & 0
                \end{array} \right)
\end{equation}
with $b$ as the strength of magnetic field. With such a
realization of those matrices, Eq.(1) becomes
\begin{equation}
  \eta \dot{\bf q} + b \hat{z} \times \dot{\bf q}
    = \nabla \psi({\bf q}) + {\bf \xi}({\bf q},t)  \; ,
\end{equation}
the very familiar form of equation of motion for a massless
particle in the presence of frictional and Lorentz forces. Here
$\hat{z}$ is the unit vector perpendicular to the state space
formed a plane by $q_1$ and $q_2$, indicating the direction of the
magnetic field $b$ with the electric charge taken to be 1. The
word ``massless" here implies that the mass of the particle is so
small it can be taken to zero in the present dynamical
consideration, and the particle dynamics is dominated by the
frictional and Lorentz forces.

The friction matrix $A$ is connected to the stochastic force ${\bf
\xi}$ by Eq.(2), which guarantees that it is semi-positive
definite and symmetric. All $T, A, \psi$ can be nonlinear
functions of the state variable ${\bf q}$ as well as the time $t$.
The numerical parameter $\epsilon$ corresponds to an effective
temperature, which can be taken to be zero to recover the
deterministic dynamics. It has been shown in section IV that if a
steady state distribution $\rho({\bf q})$ in state space exists,
\[
    \rho({\bf q}) = \frac{1}{Z} \exp
        \left(  \frac{\psi ({\bf q}) }{\epsilon } \right) \; ,
\]
a Boltzmann-Gibbs type distribution function \cite{ao2004}, same
as Eq.(6). Here $Z$ is the partition function $Z = \int d {\bf q}
\; \exp( \psi ({\bf q}) / \epsilon )$. The only use of ``mass" is
to establish such a classical equilibrium distribution physics, as
illustrated by Eq.(38). Above equation implies that for dynamics
which repeats itself indefinitely, such as limit cycle, the
potential should be the same along such trajectory.

It should be pointed that the potential function $\psi ({\bf q})$
exists from the beginning by construction. This is one of most
useful and important quantitative concepts in physics. If the
stochastic force could be set to be zero, ${\bf \xi}({\bf q}, t) =
0$, that is, the deterministic dynamics, the dynamics of this
massless particle always decreases its potential energy:
\[
 \begin{array}{lcl}
   \dot{\bf q}^{\tau} \nabla \psi ({\bf q})
    & = &  \dot{\bf q}^{\tau} [ A({\bf q}) + T ({\bf q}) ]
            \dot{\bf q}   \\
    & = &  \dot{\bf q}^{\tau} A({\bf q}) \dot{\bf q}  \\
    & \geq &  0 \; .
 \end{array}
\]
Here $ \dot{\bf q}^{\tau} \nabla \psi ({\bf q}) = \dot{\bf q}
\cdot \nabla \psi ({\bf q})$. The zero occurs only at the
invariant sets: fixed points (point attractors), limit cycles
(periodic attractors) and/or more complicated ones. Hence the
potential function has the usual meaning of Lyapunov function. We
already encountered this property in the discussion of the first
law.

To explicitly model a limit cycle, we choose following forms for
the friction matrix $A$, the anti-symmetric matrix $T$, and the
potential $\psi$, assuming the limit cycle occurs at $q_{limit\;
cycle} = 1 \; ( \; q = \sqrt{q_1^2  + q_2^2 } \; ) $:
\begin{eqnarray}
   A & = & \frac{ (q^2 -1)^2 }{ (q^2 - 1)^2 + 1 }
                           \left( \begin{array}{ll}
                                      1 & 0  \\
                                      0 & 1
                           \end{array} \right)        \; ,  \\
   T &  = & (q-1) \frac{ q^2 }{ (q^2 - 1 )^2 + 1 }
                                 \left( \begin{array}{ll}
                                          0 & -1  \\
                                         1 & 0
                                  \end{array} \right) \; ,   \\
   \psi & = & - \frac{1}{2} (q - 1)^2                 \; .
\end{eqnarray}
The potential $\psi$ given in Eq.(61) is rotational symmetric in
the state space and $|\nabla \psi | = | q-1 | $. It has a local
minimum $\psi = - 1/2$ at $q=0$, which is a cusp, and the maximum
$\psi = 0$ at $q=1$, which is a cycle in the state space. Hence
the potential takes the shape of a Mexican hat.

If the friction matrix $A$ would be zero, the dynamical
trajectory of the massless particle would move along the equal
potential contour determined by the initial condition, which would
be a cycle according to above chosen potential. In the presence of
nonzero friction matrix, this is not true. What will be our
concern is the behavior near the minimum of the potential
function: When $q$ is sufficiently close to 1, does the particle
trajectory asymptotically approach the cycle of $q=1$ and
eventually coincide with it? If the answer is positive, we have a
limit cycle dynamics. We will demonstrate below that it is indeed
possible.

For a deterministic dynamics, we can set $\epsilon = 0$ in Eq.(1)
and (2): setting the stochastic force to be zero. The dynamical
equation can then be rewritten as
\begin{equation}
      \dot{\bf q}
          = [ A({\bf q}) + T ({\bf q}) ]^{-1}
             \nabla \psi ({\bf q}) \; .
\end{equation}
With the choice of Eqs. (59,60), we have
\begin{eqnarray}
  [ A + T ]^{-1}
   &  =  & \frac{1}{\det(A + T ) }
       \left[  \frac{ (q^2 -1)^2 }{ (q^2 - 1)^2 + 1 }
        \left( \begin{array}{ll}
                 1 & 0  \\
                 0 & 1
        \end{array} \right)               \right. \nonumber  \\
  &  &  \left. - (q-1) \frac{ q^2 }{ (q^2 - 1 )^2 + 1 }  )
            \left( \begin{array}{ll}
                 0 & -1  \\
                 1 & 0
            \end{array} \right)     \right]
\end{eqnarray}
and
\begin{equation}
 \det(A+T) = \left[ \frac{(q^2 - 1)^2 }{ (q^2 - 1)^2 + 1 } \right]^2
      + \left[ \frac{ (q - 1) q^2  }{(q^2 - 1)^2 + 1 } \right] ^2 \; .
\end{equation}
Near $q=1$, we have
\begin{eqnarray}
  [ A + T ]^{-1} & =  & \frac{1}{q-1} \left[ - (1-  2  (q-1)  )
           \left( \begin{array}{ll}
                 0 & -1  \\
                 1 & 0
           \end{array} \right)  \right.  \nonumber  \\
           &   &    + \left.  4 (q-1)
            \left( \begin{array}{ll}
                 1 & 0  \\
                 0 & 1
            \end{array} \right)     + O( (q-1)^2 )  \right]  \; .
\end{eqnarray}
In terms of radial coordinate $q$ and azimuthal angle $\theta$
%
%
in the polar coordinate representation of the state space, using
the small parameter expansion given in Eq.(65) and following
Eq.(62) we have, to the order of $q-1$,
\begin{eqnarray}
   \dot{q}       & = & -  4  (q-1)  \; , \\
   \dot{\theta } & = & 1 -  2 \frac{q-1 }{q} \; .
\end{eqnarray}
The solution is
\begin{eqnarray}
  q(t)      & = & 1+\delta q_0 \exp\left\{ - 4 t \right \} \; , \\
  \theta(t) & = & \theta_0 + t
       + \frac{1}{4} \ln \frac{1+ \delta q_0
          \exp\left\{ - 4  t \right\} }
           {1+ \delta q_0  }  \nonumber  \\
    &  &  + \frac{1}{2} \ln\left(1+ \delta q_0
      \exp\left\{-  4  t \right\} \right)     \; .
\end{eqnarray}
Here $\delta q_0$ ($| \delta q_0 | << 1$) is the starting radial
position of the particle measured from $q=1$ and $\theta_0$ its
the starting azimuth angle. Indeed, the solution demonstrates the
asymptotical approaching to the cycle $q=1$, and the motion never
stops. Unstable and metastable limit cycles may be constructed in
the similar manner.

Though above construction does show that based on the physics
knowledge one can construct limit cycle with the potential, the
example of Eqs.(59-61) appears contrived. We should, however,
point out that there are several generic features in our
construction.

(i) To have an indefinite motion on a closed trajectory, because
of energy conservation, the friction, or better the friction
matrix here, must be zero.

(ii) Because of the asymptotic motion is on the equal potential
contour, a Lorentz force type must exist to keep the motion on the
contour. This means that the antisymmetric matrix should be finite
along the equal motion contour when the potential gradient is
finite. The speed of the {\it massless} particle moving along the
contour will be determined by the ratio of the strength of the
Lorentz like force to that of the gradient of the potential.

(iii) The limit cycle should be robust: Small parameter changes
should only have a small effect on the limit cycle, the (stable)
limit cycle should be at the minimum of the potential. As a
consequence, the potential gradient at the maximum is zero, which
would imply that the friction matrix must go to zero faster than
that of the potential gradient when approaching to the maximum to
avoid the potential taking singular values. This means that at the
limit cycle the dynamics is conservative.

(iv) Furthermore, to ensure the massless particle moves in the
same direction on both sides of the limit cycle, the magnetic
field should change its sign at the limit cycle. All those
features are explicitly implemented in the choice, Eqs.(59-61).

Three additional remarks are in order.

(v) For simplicity of calculation we have chosen the friction
matrix to be proportional to aunit matrix in Eq.(59). One can
check that any positive definite symmetric matrix can lead to same
conclusion, as long as its strength goes to zero in a higher order
comparing to that of the potential gradient.

(vi) Although the friction matrix $A$ is zero when approaching to
the limit cycle, $[A + T]^{-1} + [A - T ]^{-1}$ is not, to which
we will come back in subsection V.C when discussing the diffusion
matrix $D$.

(vii) It should be emphasized here that the potential has a dual
role: Its gradient is the driving force in dynamics, expressed in
Eq.(1) or Eq.(62), and it determines the final steady state
distribution, expressed in Eq.(6).

We should point out that the Mexican hat type potential has been
suggested for limit cycle previously \cite{pav}. However, to our
knowledge there is no prior discussion of the associated magnetic
and dissipative forces similar to the present and next
subsections.

\subsection{Limit Cycle in Classical Formulation}

In the above subsection we have shown that starting with potential
one can construct limit cycle. In this subsection we will show the
reverse. One may regard that the demonstration of the co-existence
of limit cycle with the potential in above subsection may be
special: The massless particle moves along the potential minimum
with both zero friction and zero transverse matrices, a rather
nice but contrived picture from physics. One may wonder that
whether or not in a typical limit cycle in dynamical systems a
potential can be constructed. We will demonstrate in this
subsection that the answer to such question is positive.

It has been suggested that a simple limit cycle in two dimensions
would take following form for its dynamical equation in a polar
coordinate representation \cite{murray}:
\begin{eqnarray}
   \dot{q }     & = &  R(q)       \; ,  \\
   \dot{\theta} & = &  \Phi (q)   \; .
\end{eqnarray}
Here the smooth functions $R,\Phi$ have properties $R(q =  1) = 0
$ is a fixed point in the radial coordinate and $\Phi (q=1) =
constant$. We point out that mathematically any shape of limit
cycle in two dimension can be deformed into a cycle and that near
this limit cycle the dynamical equation can be mapped onto above
form by a nonlinear coordinate transformation. Hence Eq.(70) and
(71) may be regarded as a representation for a generic limit cycle
in two dimension.

The comparison between Eqs.(66,67) and Eqs.(70,71) immediately
suggests what considered in section II is just such a typical
limit cycle. The construction of potential from Eq.(70) and (71)
is also immediately suggested: Given functions $R(q)$ and
$\Phi(q)$, there exist the friction matrix $S$, the potential
$\psi$, and the antisymmetric matrix $T$. In fact, we have three
independent quantities to be constructed instead of two. Such
uniqueness question has been addressed in section IV.

To construct the potential from Eq.(70) and (71), we go back to
Eq.(62). Using $[ A + T ]^{-1} = G = D + Q $ defined in Eq.(28),
Eq.(2) may be rewritten as
\begin{equation}
  \dot{\bf q} = [D({\bf q}) + Q({\bf q}) ] \nabla \psi({\bf q}) \; .
\end{equation}
Here again $D$ is a symmetric and positive definite matrix and $Q$
an antisymmetric matrix. For simplicity, we choose $D$ to be the
identify matrix,
\begin{equation}
 D = \left( \begin{array}{ll}
               1 & 0  \\
               0 & 1
             \end{array} \right) \; ,
\end{equation}
and choose
\begin{equation}
 Q = a(q) \left( \begin{array}{ll}
               0 & -1  \\
               1 & 0
             \end{array} \right) \; .
\end{equation}
Here $a(q)$ is a scalar function of $q$.

In accordance with the rotational symmetry in Eq.(70) and (71), we
also choose the potential function depending only on the radial
coordinate $q$, $\psi(q)$. With above choices, in the polar
coordinates Eq.(72) implies
\begin{eqnarray}
   \dot{q }       & = & {d \psi(q)\over{d q}}  \; ,  \\
   q \dot{\theta} & = & a(q) {d \psi(q)\over{d q}}    \; .
\end{eqnarray}
Comparing above two equations with Eq.(70) and (71) we have
\begin{eqnarray}
   {d \psi(q)\over{d q}} & = & R(q)  \; ,  \\
    b(q) & = &  q { \Psi(q) \over R(q) }    \; .
\end{eqnarray}
Thus,
\begin{eqnarray}
 A(q) + T(q)
   & = & [ D(q) + Q(q) ]^{-1} \nonumber \\
   & = & \left( \begin{array}{ll}
               1 & q { \Psi(q) \over R(q) }  \\
             - q { \Psi(q) \over R(q) } & 1
             \end{array} \right)^{-1}  \nonumber \\
    & = & { 1 \over{ 1+ \left(q { \Psi(q) \over R(q) }\right)^2}}
             \left( \begin{array}{ll}
               1 & - q { \Psi(q) \over R(q) }  \\
               q { \Psi(q) \over R(q) } & 1
             \end{array} \right)  \; .
\end{eqnarray}
This gives,
\begin{eqnarray}
 A(q)
   & = & { 1 \over{ 1+ \left(q { \Psi(q) \over R(q) }\right)^2}}
             \left( \begin{array}{ll}
               1 & 0   \\
               0 & 1
             \end{array} \right) \; ,  \\
 T(q)
   & = & { 1 \over{ 1+ \left(q { \Psi(q) \over R(q) }\right)^2}}
             \; q { \Psi(q) \over R(q) }
             \left( \begin{array}{ll}
               0 & - 1  \\
               1 & 0
             \end{array} \right)  \; .
\end{eqnarray}
Eqs.(77,80,81) are one explicit construction of potential for the
limit cycle dynamics described by Eq.(70) and (71), corresponding
to the dynamics described by Eq.(1) and (2).

When approaching to the limit cycle, $q \rightarrow 1 $, $R(q) =
O(q-1)$ and $\Phi(q) = O(1)$, and,
\begin{eqnarray}
   {d \psi(q)\over{d q}} & = & O(q-1) \; ,  \\
    T(q) & = & O(q-1)
           \left( \begin{array}{ll}
               0 & - 1  \\
               1 & 0
             \end{array} \right) \; ,    \\
    A(q) & = & O((q-1)^2)
           \left( \begin{array}{ll}
               1 & 0   \\
               0 & 1
             \end{array} \right)  \; .
\end{eqnarray}
This has the same structures as those in Eqs.(59-61). The friction
matrix $A$ indeed vanishes faster, though the diffusion matrix $D$
here always remains a constant. The dynamics eventually becomes
non-dissipative along the limit cycle from the point of view of
Eq.(1). This completes the discussion of construction of potential
for a limit cycle in two dimension.

For an arbitrary limit cycle, the construction of potential from
the classical formulation will start from Eq.(24) and (27), which
are first order partial differential equations. The behavior near
a limit cycle should resemble that described by Eqs.(82-84). Such
behavior provides the necessary boundary condition to solve
Eq.(24) and (27). Only in some extremal cases the solving of
partial differential equations can be turned into the solving of
algebraic equation, as showed in this subsection.

Nevertheless, we should point out an interesting feature. On one
side it is known from the theory of dynamical systems that a limit
cycle is robust \cite{guckenheimer}. Any small parameter change in
the equation would not lead to its disappearing.  On the other
side, from the demonstration in section II which is based on the
understanding from physics, the existence of the limit cycle is a
result of a very delicate balance between all dynamical elements:
The friction matrix, the gradient of potential function, and the
antisymmetric matrix. They are all zero at the limit cycle, and
when approaching to the limit cycle, the friction matrix vanishes
faster. How this paradoxical feature would play a role in our
better understanding limit cycle and its control will be an
interesting problem for further exploration.

\subsection{ Diffusion {\it vs} Ascendency (Friction) }

There are two mathematical subtleties. First, it is known that
even with a limit cycle the dynamics is dissipative, reflecting by
the fact that in general $\nabla \cdot {\bf f} \neq 0 $, where
${\bf f}$ is defined by Eq.(14). This is also expressed by the
fact that the so-called diffusion matrix, $D$ defined in Eq.(15),
is finite even at the limit cycle. Because it is dissipative, it
would be difficult to conceive a constant potential (or a Lyapunov
function) along the limit cycle on which the dynamics repeats
itself indefinitely. The delicate point is that, as shown in our
above demonstration, that the friction matrix $A$ is zero along
the limit cycle does not implies the diffusion matrix $D$ is zero.
In fact, it is finite according to Eq.(22) and (27):
\begin{eqnarray}
   D & = & \frac{1}{2} [ G + G^{\tau} ] \nonumber \\
     & = & \frac{1}{2} [ (A + T)^{-1} + (A - T)^{-1} ]  \;  .
\end{eqnarray}
An explicit verification can be obtained from Eq.(65): $D = 4$
while $A = 0$ along the limit cycle. An important and direct
implication of present construction is that the statement, that
there is no evolutionary potential (or Lyapunov function) in limit
cycle in \cite{rice}, is not valid.

We come to the second subtlety. For completely deterministic
dynamics Lyapunov function cannot be uniquely defined: If one
finds one Lyaponov function, one finds many. This is illustrated
by the present construction that additional information from the
noise is needed to make the construction unique: different
diffusion matrix would lead to different potential function. We
already encountered this issue when construction potential from
Eq.(70) and (71), without the specification of noise. However,
this freedom may provide a method to select the best suitable
Lyapunov function or potential function to one's problem by
choosing appropriate form of the diffusion matrix.

It is worthwhile to point out a simple mathematical fact that the
typical gradient systems in dynamical systems theory corresponds
to the zero transverse matrix, $T=0$, in the present construction.
It is the case when the detailed balance condition is satisfied.
No limit cycle is possible in this case. In dynamics described by
gradient systems the trajectory could follow the most rapid
descendant path along the landscape defined by the potential. In
this case it is easy to identify the potential as the landscape
function.  For a general dynamics where the transverse matrix is
not zero, the trajectory would not follow the most rapid
descendant route along the potential, as expressed by Eq.(1) or
Eq.(62). Nevertheless, the meaning of the potential remains the
same as that in gradient systems: driving the dynamics and
determining the final steady state distribution.

\section{F-Theorem: Further Explorations}

It is now clear that the Fisher's original formulation of the
fundamental theorem of natural selection is in the domain of
classical formulation. To illustrate this point, let us consider
the situation that the detailed balance condition, Eq.(48), is
satisfied. From Eq.(14) and (15), we have
\begin{equation}
  \dot{\bf q} = D({\bf q}) {\nabla} \psi({\bf q}) +
                    {\bf \zeta}( {\bf q}, t) \; .
\end{equation}
Thus, the rate of increase is indeed proportional to the variance
$D$. This result can be made most explicit in one dimension. In
one dimension, the transverse matrix is zero by definition. Near a
potential peak, taking it as $q_{peak}=0$,
\begin{equation}
 \psi(q) = \psi(0) - \frac{1}{2} U q^2 + O( q^3 ) \; ,
\end{equation}
and
\begin{equation}
 \nabla \psi = - U q \; .
\end{equation}
Here in one dimension $U$ is a positive constant. If we also drop
the noise term for simplicity, from Eq.(86) and (87) we have
\begin{eqnarray}
  \frac{ \frac{ d(\psi(q) - \psi(0))}{dt} }{|\psi(q) - \psi(0)|}
        & = & - 2 \frac{ \dot{q} }{q}  \nonumber  \\
        & = & 2 D(0) U  \; .
\end{eqnarray}
The rate of increase in evolutionary potential is indeed
proportional to the variance $D$ in the classical formulation,
with a suitable choice of unit as indicated by the positive
constant $U$. This corresponds precisely to Fisher's original
statement.

It is interesting to note that under those conditions the rate of
increase in evolutionary potential is proportional to the inverse
of the variance $A$ ($D = 1/A$ under detailed balance condition)
in the present formulation, Eq.(2). There is nevertheless no
contradiction here. In the present formulation the meaning is
taken from the fluctuation-dissipation theorem: the return to the
evolutionary potential peak and the fluctuation from the peak are
governed by the same process. This meaning is also implied in
Fisher's fundamental theorem of natural selection. No other direct
connection between rate of increase in evolutionary potential and
the variance $A$ has been stated in the present formulation,
except what can be obtained from Eq.(1) and (2).

Such demonstration suggests that Fisher's original formulation is
only exact near the peak of evolutionary potential and under the
condition of detailed balance. While the requirement to be at the
vicinity of the peak does not appear to a major barrier to
generalize Fisher's original statement to nonlinear case of far
from the peak, the requirement of detailed balance is too strong.
The latter can be violated even near the peak, if transverse
matrix $T$ is no longer zero. It causes the breakdown of detailed
balance. In the absence of detailed balance condition, we no
longer have the simple relation $D = 1/A $, as implied by Eq.(21).
It is known that the breakdown of detailed balance is common in
population genetics. This would result in the situation that the
variance, the diffusion constant $D$, can only partially determine
the rate of increase in evolutionary potential. We further
illustrate this point below.

Near a potential peak, taken to be ${\bf q}_{peak} = 0$, in
dimension higher than one, we may approximate the diffusion matrix
$D$ by a constant matrix. The force ${\bf f}$ can be linearized:
\begin{equation}
  {\bf f}({\bf q}) = S {\bf q} \; .
\end{equation}
Here $S$ is the selection matrix defined by Eq.(18) in subsection
IV.A and is a constant matrix in the linear case. Similarly, the
Wright evolutionary potential can be expressed by a constant
potential matrix $U$:
\begin{equation}
 \psi({\bf q}) = \psi(0) -
    \frac{1}{2} {\bf q}^{\tau}U{\bf q} \; ,
\end{equation}
and the ascendant matrix $A$ is also a constant matrix. Here $U,
A, D$ are symmetric and positive definite while $S$ is asymmetric.

Given $D$ and $S$ in the classical formulations, $U$, $A$  and $T$
can be uniquely determined \cite{kat}. In general they are
complicated function of $D$ and $S$. Formally, we still have a
simple formula similar to Eq.(89):
\begin{equation}
  \frac{ \frac{ d(\psi(q) - \psi(0))}{dt} }{|\psi(q) - \psi(0)|}
         =  2 \frac{{ \bf q}^{\tau}U(D,S) D U(D,S) {\bf q} }
                   {{\bf q}^{\tau}U(D,S){\bf q}  }  \; .
\end{equation}
There is, however, no simple interpretation of this formula in the
absence of detailed balance condition. In the limit there is a
strong asymmetric dynamics, that is, if the transverse matrix $T$
is much larger than the ascendant matrix $A$: $| \det(T) | >>
\det(A)$, to the first order in ascendant matrix we have
\begin{equation}
  \frac{ \frac{ d(\psi(q) - \psi(0)) }{dt} }{|\psi(q) - \psi(0)|}
         =  2 \frac{ ( T^{-1} U {\bf q})^{\tau} A (T^{-1} U {\bf q}) }
                   { {\bf q}^{\tau}U{\bf q}  }  \; ,
\end{equation}
which is proportional to the variance $A$ in the present
formulation, because $A,T,U$ are independent quantities. Again,
there is no contradiction here: it is still that the return to the
evolutionary potential peak and the fluctuation from the peak are
governed by the same process.

Initially there may be a large deviation from the peak and the
fluctuation may be relative small. As time progresses, the system
gets closer and closer to the peak, and the deviation $|\psi(0) -
\psi({\bf q})|$ becomes smaller and smaller. Eventually we can no
longer differentiate the deviation from the fluctuation. Both the
return to the peak and the fluctuation have to be considered
simultaneously. Statistically, the distribution of state variable
averaged over time is then given by the distribution in the form
of Eq.(8), well-known in population genetics \cite{kisdi}.  This
suggests that the staying in the peak is a dynamical process: the
balance of fluctuation from the peak and the tendency to return.
This is similar to van Valen's Red Queen hypothesis
\cite{vanvalen}: in order to stay at the same position the motion
is required.

The partial increase in evolutionary potential by variance has
already been observed in population genetics \cite{ewens}. In
section V we have discussed an extreme case: while the variance
$D$ in the classical formulation is finite, there is no change at
all in the evolutionary potential when the system moves along a
limit cycle. Such a consideration suggests that Fisher's original
formulation of the fundamental theorem of natural selection is
indeed not general enough. The present formulation can deal with
those situations in a consistent manner. Above analysis also
suggests that the conservative dynamics represented by the
transverse matrix $T$ must be an integral part of general
evolutionary dynamics.

We have now demonstrated that the contexts of the fundamental
theorem of natural selection and the fluctuation-dissipation
theorem are the same: the connection between the fluctuation and
the ability to search for peaks. Nevertheless, there is still an
imperfection in those formulations: A potential peak, or, in
general the evolutionary potential, is implicitly assumed in their
statements. On the other hand, in the present formulation of the
F-Theorem there is no reference to the evolutionary potential. The
connection between the fluctuation and the ability to search in
state space is still kept.  This suggests that the F-Theorem is a
generalization of the fundamental theorem of natural selection,
making it applicable to the cases where there is no change in
evolutionary potential, such as the case of limit cycle dynamics
in section V or the case of speciation in the holey adaptive
landscape \cite{gavrilets}.

\section{ Discussions }

Before further discussion of the implications of the present
mathematical formulation, it should be kept in mind that the three
laws must be regarded as a first attempt to formulate
mathematically what the evolutionary dynamics might be. Further
generalizations or extensions are needed. For example,
Gaussian-white noise assumption may well be the first step towards
a complete modelling of stochasticity in evolution. We have not
discussed the nonwhite noises, though we believe the presentation
formulate gives us a good starting point to do so. In this
connection we point out that the famous $1/f$ noise ($s=0$ as
formulated in Ref.[\cite{leggett}]) may be constructed from white
noise ($s=1$ as formulated in Ref.[\cite{leggett}]) \cite{at94}.
The present discussion has been confined to stochastic
differential equations and related Fokker-Planck equations. We
believe generalization to stochastic partial differential
equations may be needed for a complete discussion of speciation.
In addition, in the present review the focus is dynamics, not the
structure of each dynamical element. Those dynamics elements may
take different forms at different levels of description, an
important subject completely outside the present scope.

Having made those qualifying observations, we nevertheless remark
that although the proposed three laws for evolutionary dynamics
are based on the continuous representation in terms of time and
population, it is possible that main features discussed in the
present review should survive in more general cases: the ultimate
selection by Wright evolutionary potential, the adaptive nature of
ascendant matrix, the conservative dynamics represented by the
antisymmetric matrix, and the presence of stochastic drive. We
have emphasized the intrinsic stochastic nature of evolutionary
process, not those due to imperfections from measurement and/or
observation.

\subsection{ Consistency and Compactness }

In the present attempt to unify approaches from biological and
physical sciences, the possible existence of general laws such as
expressed by Eq.(1) and (2) in a compact and consistent manner
should not be too surprising.
One has to consider two important principles which have been rigorously
validated:

(i) Simple equations can generate extremely complicated patterns
and phenomena \cite{may,murray,guckenheimer};

(ii) Each level of description has its own laws which cannot be
derived in a naive reductionist manner
\cite{anderson,goldenfeld,batterman}. The connections between
levels are asymptotic and emerging phenomena frequently occur at
higher levels.

There are several quantitative advantages in the present
formulation of evolutionary theory. With the Wright evolutionary
potential defined as in Eq.(1), an independent way for its
calculation has been obtained. It adds more predictive power to
the evolutionary dynamics.

As expressed by Eq.(2) and discussed in section 2.4, Fisher's
fundamental theorem of natural selection (the F-Theorem) becomes
transparent and indispensable in the present formulation of
evolutionary dynamics. This may provide a much-needed step to
better understand Fisher's great insight.

Combining both Fisher's and Wright's insights, the Wright
evolutionary potential and the F-Theorem provide a quantitative
measure to discuss robustness, stability, and the speciation.
Eq.(9) is such an example: both variance and the evolutionary
potential are needed to obtain the formula. What has been new in
the present article is the glue to put both Fisher's and Wright's
insights together: Evolutionary dynamics has a conserved part in
which the Wright evolutionary dynamics would not change. This
conservative part of dynamics is represented by the transverse
matrix in the present article, and appears to have been overlooked
in literature.

It is interesting to point out the remarkable similarity between
the adaptive landscape of Wright \cite{wright} and the
developmental landscape of Waddington \cite{waddington}. In fact,
there is a remarkable similarity between the adaptive speciation
\cite{ddmt} and development biology \cite{waddington}: Their
landscapes are identical. The present mathematical formulation
should be able to deal with both cases. An example of the gene
regulatory network in phage $\lambda$ has already been studied
successfully by Zhu {\it et al.} \cite{zhu2004a,zhu2004b}. This
suggests a unification between genetics and developmental biology.

\subsection{Usefulness}

However, one may wonder about the need to use Eq.(1) and (2)
instead of the classical formulation of  Eq.(14) and (15) or
Eq.(39): After all their equivalence has been demonstrated above.
Here we offer three reasons that Eq.(1) and (2) can be useful:

(i) Quantities presented in Eq.(1) can be directly related to
experimental observation. For example, Eq.(6) gives a direct
connection between the Wright evolutionary potential and the
population density in steady state. By observing the dynamical
behaviors, information on the ascendant and transverse matrices
can be obtained. Also, Eq.(9) can relate stability to the Wright
evolutionary potential. This direct contact with experimental data
is an indication of the autonomy of the focused level of
description.

(ii) Eq.(14) and (39) lack the visualizing ability for the global
dynamics behavior. For example, in a nonlinear dynamics with
multiple local maxima, it is not clear from Eq.(14) and (15) which
maximum is the highest and how easy it might be to move from one
maximum to another. One could find this answer by a direct real
time calculation. But this is usually computationally demanding,
if not impossible. For deterministic dynamics it is impossible.
Instead, the Wright evolutionary potential gives a quantitative
and visualizable answer to such inquiry.

(iii) Eq.(1) and (2) give an alternative modelling of evolutionary
dynamics, which can be advantageous in certain situations. For
example, the direct use of evolutionary potential (fitness
function by Stewart) in Stewart's modelling \cite{stewart} makes
the symmetry-breaking idea in speciation very transparent from
statistical physics' point of view.

\subsection{Biology and Physics}

Finally, we point out an interesting mutual reduction loop between
biology and physics.

The present formulation suggests that the laws as expressed by
Eq.(1) and (2) are stochastic and dissipative. They represent the
generic features of the dynamics in biology. For example, the
mathematical structure used in the present review was first
suggested in our study of the outstanding robustness puzzle in a
genetic switch \cite{zhu2004a,zhu2004b}, after a long search for a
way to quantify robustness and stability. On the other hand, the
dynamical law of physics is deterministic and conservative, as
manifested in the classical Newtonian dynamics. In the discussion
of the first law of Darwinian dynamics we remarked that one may
regard the Newtonian dynamics as a special case of the first law
when the nonconservative force represented by the ascendant matrix
is zero. Since the first law is a special case of the second law,
the Newtonian conservative dynamics may be further regarded as a
special case of the present second law, hence a special case of
the Darwinian dynamics. The opposite statement also exists: Under
an appropriate condition of treating the dissipative dynamics as a
subdynamics, a typical approach to model open systems in physics,
equations in the form of Eq.(1) and (2) can be derived from the
Newtonian dynamics \cite{leggett,at,az,gbg,evans}. Therefore, the
Darwinian dynamics may also be regarded as a special case of the
Newtonian dynamics.

Darwinian dynamics places an emphasis on the statistics while the
Newtonian dynamics on the determinism. Which one would be more
appropriate and fundamental to describe Nature is both a
philosophical and a serious scientific research topic. The
approach from the dissipative dynamics side can be found the work
of Prigogine school \cite{prigogine} and the approach from the
conservative dynamics side can be found, for example, in
Ref.[\cite{evans}] and [\cite{gbg}]. It appears suitable to end
this subsection with a quotation from Einstein who had thought
deeply into this issue \cite{einstein1950}:

``{\it Concerning the question of Statistics against Determinism,
this is the way it appears: From the point of view of immediate
experience there is no such thing as exact determinism. Here there
is no disagreement. The question is whether or not the theoretical
description of nature must be deterministic. Beyond that, the
question is whether or not there exists generally a conceptual
image of reality (for isolated case), an image which is in
principle completely except from statistics. Only on this subject
do opinions differ.}"

\section{ Further Notes on Literature }

While it is impossible to review all theoretical literature on
Darwinian dynamics in the present article, in this section we
discuss a few selected works related to the present formulation.
We hope that it may be helpful to readers for further pursuing the
subject.

One of most extensive discussions on Darwinian evolutionary
dynamics from biological point of view is provided by Gould
\cite{gould}.  He held a view against a more quantitative
formulation, calling it the hardening of Darwinian theory. In
particular, the first three exceptions of Fisher's fundamental
theorem of natural selection were summarized in his book as the
exceptions of contingency, individuality, and interaction. Those
exceptions, according to Gould, simply excluded Fisher's
fundamental theorem of natural selection from biology.  In the
light of present formulation, while Gould's critiques of Fisher is
correct, it is nevertheless based on wrong formulation and
improper analogy. For example, as discussed in subsection IV.D,
the fitness should not be interpreted as the entropy in physics,
and the proper analogy of Fisher's fundamental theorem of natural
selection should be the fluctuation-dissipation theorem in
physics, formulated as the F-Theorem in the present article. With
such re-interpretation, the Gould's critique disappears, because
all Fisher's five exceptions disappear. Nevertheless, Gould's book
still serves as one of the best biological sources to understand
evolution.

On a conceptual level, the controversies and key concepts of
Darwinian evolutionary theory have been concisely summarized by a
recent book by Mayr \cite{mayr2004}. It is a must read book. His
summary of Darwinian dynamics that ``the basic Darwinian formula -
evolution is a result of genetic variation and its ordering
through elimination and selection - is sufficiently comprehensive
cope with all natural eventualities" is indeed a most proper
verbal presentation of the present second law at genetic level. It
is, however, puzzling to note that, whether accidental or
intensional, there is a de-emphasis on the quantitative aspects of
population genetics in Mayr's book. It is hard to conceive the
presentation of the modern synthesis of Darwinian theory without a
detailed discussion on Wright and Fisher, particularly on the
fundamental theorem of natural selection.

In addition to the summary of mathematical results since Fisher
and Wright on population genetics by Ewens \cite{ewens}, the
formulation of Darwinian dynamics in the line of classical
formations has been presented in detail by Michod \cite{michod}.
Both the Wright's evolutionary landscape and Fisher's fundamental
theorem of natural selection occupy an prominent place in the
book. Some 28 different usages of the term {\it fitness} are
compiled in his Appendix B, which may be regarded as a support to
use the term Wright evolutionary potential in the present review,
instead of the loaded term {\it fitness}. Also interesting is his
emphasis on the cooperation and interaction, which may shed a
light on the controversy regarding to the Wright's shift balance
theory and related speciation problem \cite{goodnight,pigliucci}.

The discussion by Turchin \cite{turchin} on debate of the
existence of general laws in biology should be interesting.
Concrete examples in ecological dynamics have been discussed in
detail in his book.  Nevertheless, it seems that Turchin has not
emphasized the role played by stochasticity in general laws,
though he does emphasize the need to use statistics in data
analysis. Much of his detailed discussions is deterministic in
nature and hence may be viewed as the unfolding of the first law
discussed in the present article. For example, his first principle
of exponential growth (decrease) may be regarded as a special case
of the stable fixed point at infinity (zero). A discussion of
ecological dynamics from the adaptive landscape perspective can be
found in Schluter's book \cite{schluter}.

One of the best discussions on the classical formulation of
evolutionary dynamics in a general setting is certainly by Nicolis
and Prigogine \cite{prigogine}. A parallel and stimulating
approach can be found in the exposition of synergetics
\cite{haken}. One of the best discussions on the
fluctuation-dissipation theorem in physics can be found in
Ref.[\cite{ymr}]. We should also mention that the concept of
evolutionary potential has been elegantly employed by Stewart
\cite{stewart} in his discussion of speciation and by many other
researchers \cite{ddmt} in the deterministic study of adaptive
speciation.

\section{ Conclusions }

We have postulated three laws to mathematically describe the
evolutionary dynamics: the law of Aristotle, the law of Darwin,
and the law of Hierarchy. The first law, the law of Atistotle,
emphasizes the deterministic aspect of the evolutionary dynamics.
This law defines the reference point for discussion. The most
fundamental equation, the second law or the law of Darwin, has
been expressed in a unique form of stochastic differential
equation. Four dynamical elements have been introduced into the
present formulation: the ascendant matrix, the transverse matrix,
the Wright evolutionary potential, and the stochastic drive. The
final and ultimate selection is represented by the Wright
evolutionary potential which determines the steady state
distribution. At any given time, the instant selection is
determined by all four dynamical elements encoded in the second
law, which would not necessarily increase the Wright evolutionary
potential. Both the tendencies for adaptation and for optimization
are encoded in the F-Theorem. The hierarchical nature of
biological description is encoded in the third law whose precise
mathematical procedure depends the specific problem in
consideration. This is illustrated by the existence of several
integrations of stochastic differential equations. Table I gives a
summary of the present formulation.

We have also demonstrated that present three laws are consistent
with classical approaches in evolutionary biology, but appear more
suitable to discuss stability and other phenomena quantitatively.
Various important results, such as Fisher's fundamental theorem of
natural selection and Wright's adaptive landscape, as well as the
developmental landscape, are unified in the present formulation in
a natural manner. The inconsistent understandings of those two
central concepts of Fisher and Wright, raised either from the
original vague verbal statements or from intrinsical mathematical
difficulties, have been discussed and clarified. The main goal of
the present formulation is to resolve the outstanding and
historical problems rather than to complicate life. Indeed, it
appears to form a consistent and quantitative foundation for
further discussion of the Darwinian dynamics.

{\ }

{\bf Table I: Laws of Darwinian dynamics and the F-Theorem }

\noindent
The central dynamical equation is the Second Law
supplemented by the F-Theorem.  \\
$\psi$: Wright evolutionary potential, ultimate selection \\
$T$: antisymmetric transverse matrix, conservative dynamics \\
$A$: ascendant matrix, adaptive dynamics \\
$\xi$: stochastic drive, random \\
${\bf q}$: state variable vector

{\ }

\noindent
\begin{tabular}{|c|c|c|c|}
  \hline
    & mathematical expressions & alternative names &  comments   \\
  \hline
 First Law                     &
  $  \{ {\bf q}  \}
       \rightarrow  \{ {\bf q}_{attractor} \} $  &
    law of Aristotle           &
   determinism                 \\
 Second Law                    &
  $ [ A({\bf q}) + T ({\bf q}) ] \dot{\bf q}
     = \nabla \psi ({\bf q}) + {\bf \xi}({\bf q}, t) $ &
    law of Darwin              &
       stochastic dynamics      \\
 F-Theorem                     &
  $ \langle {\bf \xi}({\bf q}, t)
          {\bf \xi}^{\tau} ({\bf q}, t') \rangle
    = 2 A({\bf q})  \; \delta(t-t') $   &
    Fisher's FTNS              &
    adaptation and optimization    \\
 Third Law                     &
     $m$ $\rightarrow 0 $     &
    law of hierarchy           &
    multiple time scales   \\
  \hline
\end{tabular}

{\ }

{\ }

\noindent{\bf Acknowledgements:} The critical comments of W.J.
Ewens, J. Felsenstein and D. Waxman on population genetics are
appreciated. I am grateful to helpful discussions with C.T.
Bergstrom, L. Yin,
 X.-M. Zhu, Z. Chen, G. Kosaly, C. Kwon, H. Qian, D. Galas, and
 D.J. Thouless for various aspects of the underlying mathematics. I
 am also indebted to J. Felsenstein for a critical reading of the manuscript.
 The encouragement and suggestion from the editor and the extensive
 and constructive comments on both science and presentation from
 two anomalous reviewers are gratefully acknowledged.
This work was supported in part by a USA NIH grant
under HG002894.

{\ }

\end{document}